\documentclass[11pt]{article}
\usepackage{amssymb}

\usepackage[totalwidth=460truept,totalheight=600truept]{geometry}
\usepackage{latexsym,amssymb,amsmath,graphicx,accents,eucal,slashed,subfigure}
\usepackage{dsfont}
\usepackage[T1]{fontenc}

\def\theequation{\arabic{section}.\arabic{equation}}

\renewcommand{\theequation}{\thesection.\arabic{equation}}
\global\arraycolsep=1truept

\numberwithin{equation}{section}
\renewcommand{\theequation}{\arabic{section}.\arabic{equation}}

\begin{document}

\phantom{C}

\vskip 1.4truecm

\begin{center}
{\huge \textbf{Distance Between Quantum Field Theories}}

\vskip .3truecm

{\huge \textbf{As A Measure Of Lorentz Violation }}

\vskip 1truecm

\textsl{Damiano Anselmi}$^{a,b}$\textsl{\ and Dario Buttazzo}$^{c}$

\vskip .2truecm

$^{a}$\textit{Institute of High Energy Physics, Chinese Academy of Sciences,}

\textit{19 (B) Yuquanlu, Shijingshanqu, Beijing 100049, China,}

\vskip .2truecm

$^{b}$\textit{Dipartimento di Fisica ``Enrico Fermi'', Universit\`{a} di
Pisa, }

\textit{Largo B. Pontecorvo 3, I-56127 Pisa, Italy,}

\vskip .2truecm

$^{c}$\textit{Scuola Normale Superiore, }

\textit{Piazza dei Cavalieri 7, I-56126 Pisa, Italy}

\vskip .2truecm

damiano.anselmi@df.unipi.it, dario.buttazzo@sns.it

\vskip 1.5truecm

\textbf{Abstract}
\end{center}

\medskip

{\small We study the distance between symmetry-violating quantum field
theories and the surface of symmetric theories. We use this notion to
quantify how precise Lorentz symmetry is today, according to experimental
data. The metric in parameter space is defined \textit{\`{a} la}
Zamolodchikov, from the two-point function of the Lagrangian perturbation.
The distance is obtained minimizing the length of paths connecting the
Lorentz-violating theory to the Lorentz surface. This definition depends on
the Lagrangian used to formulate the theory, including total derivatives and
the choice of coordinate frame. We eliminate such dependencies minimizing
with respect to them. We derive a number of general formulas and evaluate
the distance in the CPT-invariant, QED subsectors of the Standard Model
Extension (SME) and the renormalizable high-energy-Lorentz-violating
Standard Model. We study the properties of the distance and address a number
of applications.}

\vskip 1truecm

\vfill\eject

\section{Introduction}

\setcounter{equation}{0}

Theories that explicitly violate Lorentz symmetry contain a large number of
independent parameters. For various purposes, it can be useful to collect
them into a single quantity that measures the ``amount of violation''. In
this paper we propose to achieve this goal computing the distance between
the Lorentz-violating theory and the Lorentz-symmetric surface. More
generally, we study the distance between symmetry-violating theories and the
symmetric surface. We investigate the properties of the distance and
calculate it in quantum electrodynamics, using the known experimental bounds
on the parameters of the Lorentz violation \cite{datatables}.

When a symmetry is broken spontaneously, there typically exists one
parameter $\mu $, such as the vacuum expectation value of a scalar field,
that quantifies the amount of violation. At $\mu =0$ the symmetry is
restored, while at $\mu \neq 0$ $\mu $ can be viewed as a measure of the
violation. When a symmetry is violated explicitly, instead, the parameters
of the violation are typically numerous. The distance between a
symmetry-violating theory and the surface of symmetric theories is a single
quantity that collects all parameters of the violation, and vanishes if and
only if all of them vanish. When it is infinitesimal, it is a relatively
simple positive-definite quadratic form. It can be useful for several
purposes. First, it assigns a relative weight to each parameter of the
violation. Some parameters may be more or less significant than others, in
some limits or particular situations. This kind of knowledge may be useful
to guide the experimental search. The notion of distance can also address
the search for a unifying principle behind the violation, or a more
fundamental completion of the theory. Finally, it can be useful also in
effective field theory, because at low-energies a spontaneously broken
symmetry may look indistinguishable from an explicitly broken symmetry.

We assume that a theory $\tau _{\mathcal{L}}$ is defined by a Lagrangian $%
\mathcal{L}$ and a quantization procedure, such as the functional integral.
We start from the metric in parameter space, defined \textit{\`{a} la}
Zamolodchikov \cite{zamo}, from the two-point function of the infinitesimal
Lagrangian perturbation. The metric defines the infinitesimal distance $%
\mathrm{d}\ell $ between two theories $\tau _{\mathcal{L}}$ in the usual
way. Integrating $\mathrm{d}\ell $ along a path $\gamma $ gives the length
of the path. The (finite) distance between two theories $\tau _{\mathcal{L}}$
is then the length of the shortest path connecting them. This definition is
renormalization-group (RG) invariant and satisfies the axioms of a distance,
but has some unusual features. For example, it strictly depends on the
Lagrangian used to formulate the theory. Total derivatives and coordinate
reparametrizations, among the other things, do affect the distance. A way to
eliminate such dependencies is to suitably minimize with respect to
equivalent Lagrangian formulations of the same theory.

The distance between a theory $\tau _{\mathcal{L}}$ and a surface $S$ of
theories is the length of the shortest path connecting $\tau _{\mathcal{L}}$
to $S$. For theories infinitesimally close to $S$ it is sufficient to
calculate the normal vector, for which we give a simple formula in terms of
a ``reduced metric'' that incorporates the effects of the minimization.

Variations of parameters that do not move away from the symmetric surface
will be called \textit{tangent displacements}. Minimizing with respect to
tangent displacements is necessary to find the closest point on the surface.
When we consider variations of parameters introduced by reparametrizations
of fields and coordinates, we speak of \textit{%
reparametrization-displacements}. Minimizing with respect to
reparametrization-displacements is a way to eliminate the ambiguities
associated with the formulation of the theory. Alternatively, the
ambiguities can be removed with conventional prescriptions. In some cases
Lorentz-violating theories remove some of these ambiguities automatically,
because they define a preferred reference frame.

We evaluate the distance in Lorentz-violating quantum electrodynamics. At
low-energies, we consider the CPT-invariant sector of the
Colladay-Kostelecky minimal Standard Model extension of \cite{colladay}, and
use the data tables of ref. \cite{datatables}. The minimizations with
respect to tangent- and reparametrization-displacements give constraints
that coincide with the conventions used in the literature.

Later we extend the calculation to the QED subsector of the
Lorentz-violating Standard Model of ref.s \cite{lvsm,noH}. It includes
higher-dimensional operators, but it is still renormalizable by weighted
power counting \cite{halat}. In this case we use the bounds of ref. \cite
{mewes}.

We also derive formulas for the most general CPT-invariant marginal
Lorentz-violating deformations of free relativistic fields and study the
conditions obtained minimizing with respect to tangent- and
reparametrization-displacements.

Quantum electrodynamics is weakly coupled at all energies we are interested
in, and most experimental bounds on the parameters of the Lorentz violation
are very small. For this reason, one-loop results and the free-field limit
are sufficient to calculate the distance to a first approximation. In the
marginal sector, the relative weights assigned by the distance to
Lorentz-violating parameters are simple numerical factors of order 1.
Indeed, no very large or small numbers can be generated by one-loop diagrams
involving marginal operators. This means that all marginal parameters are on
an equal footing.

On the other hand, when higher-derivative operators are included, the
relative weights can be larger numbers, and may lower the scale at which the
effects of the Lorentz violation become important. The distance can be
useful as a guiding quantity to identify which parameters at which energies
are more significant to search for signs of Lorentz violation.

As mentioned earlier, the distance may depend on total derivatives added to
the Lagrangian. In general, every dependence on unwanted parameters can be
eliminated minimizing with respect to them. Then, however, formulas become
considerably involved. Sometimes it may be convenient to fix suitable
prescriptions instead of minimizing. We study several options, and show that
the qualitative features of the distance are unaffected by these choices.

The paper is organized as follows. In section 2 we define the metric and the
distance in parameter space. In section 3 we study the distance between a
Lorentz-violating theory and the Lorentz surface, and derive some useful
formulas, in particular when the distance is infinitesimal. In section 4 we
study simple examples, such as infinitesimal and finite distances among
massive and massless fields, and the case of spontaneously broken gauge
symmetries. In section 5 we derive formulas for the most general marginal
CPT-invariant Lorentz-violating deformations of free massless relativistic
fields. In section 6 we evaluate the distance in Lorentz-violating quantum
electrodynamics. We first focus on the low-energy limit and then include
higher-derivative operators, in particular those predicted by the
Lorentz-violating Standard Model of ref.s \cite{lvsm,noH}. In section 7 we
study the dependence of the distance on total derivatives and coordinate
reparametrizations. In the appendix we show that the distance is RG
invariant.

We need to use both Minkowskian and Euclidean notations. All parameters used
in our formulas are Minkowskian, which is convenient to emphasize the
positive-definiteness of the distance, and make contact with existing
parametrizations. The Euclidean notation is convenient to define and
calculate the distance.

\section{Metric and distance in parameter space}

\setcounter{equation}{0}

In this section we define the metric in parameter space, the length of a
path, and the distance between theories and from a theory to a surface of
theories.

We assume that a quantum field theory $\tau $ is described by a Lagrangian $%
\mathcal{L}$ and a quantization procedure. For definiteness, we may choose
the functional-integral approach and the dimensional-regularization
technique. A Lagrangian uniquely defines a theory, but the same theory can
be described by different Lagrangians. We call $\mathcal{L}$-theory the
theory $\tau $ as it is defined by the Lagrangian $\mathcal{L}$, and denote
it with $\tau _{\mathcal{L}}$. We must first define the distance between two 
$\mathcal{L}$-theories, then the distance between two theories.

Given an $\mathcal{L}$-theory with Lagrangian $\mathcal{L}(\lambda )$ and
couplings $\lambda ^{I}$, consider a small perturbation 
\[
\mathcal{L}+\mathrm{d}\mathcal{L}=\mathcal{L}+\sum_{I}\mathrm{d}\lambda ^{I}%
\mathcal{O}_{I}=\mathcal{L}(\lambda +\mathrm{d}\lambda ), 
\]
where the $\mathcal{O}_{I}$s are local operators. The infinitesimal squared
distance between $\tau _{\mathcal{L}}$ and $\tau _{\mathcal{L}+\mathrm{d}%
\mathcal{L}}$ at energy $E=1/\hat{x}$ is defined as 
\begin{eqnarray}
\mathrm{d}\ell ^{2} &=&2\pi ^{4}\hat{x}^{8}\langle \mathrm{d}\mathcal{L}%
\left( \frac{x}{2}\right) \hspace{0.01in}\hspace{0.01in}\mathrm{d}\mathcal{L}%
^{\theta }\left( \frac{x^{\theta }}{2}\right) \rangle =2\pi ^{4}\hat{x}%
^{8}\langle \mathrm{d}\mathcal{L}(\hat{x}_{v})\hspace{0.01in}\hspace{0.01in}%
\mathrm{d}\mathcal{L}^{\theta }(0)\rangle  \nonumber \\
&=&2\pi ^{4}\hat{x}^{8}\sum_{IJ}\mathrm{d}\lambda ^{I}\langle \mathcal{O}%
_{I}(\hat{x}_{v})\hspace{0.01in}\hspace{0.01in}\mathcal{O}_{J}^{\theta
}(0)\rangle \mathrm{d}\lambda ^{J*}\equiv \sum_{IJ}\mathrm{d}\lambda
^{I}g_{IJ}(\hat{x},\lambda )\mathrm{d}\lambda ^{J\hspace{0.01in}*},
\label{distance}
\end{eqnarray}
where $\langle \cdots \rangle $ is the expectation value on the vacuum state
of the unperturbed theory $\mathcal{L}$ in Euclidean space, $x=(\hat{x},\bar{%
x})$, $\hat{x}>0$, and $\hat{x}_{v}$ denotes the four-vector $(\hat{x},%
\mathbf{0})$. Hats denote time components and bars denote space components.
In both Minkowski and Euclidean spaces time components will also be denoted
with the index 0. Finally, $\theta $ denotes the operation of time
reflection. On coordinates $x$ it acts as $x^{\theta }=(-\hat{x},\bar{x})$.
On operators in Euclidean space it acts antilinearly and generates a factor $%
-1$ for every time index: 
\[
\mathcal{O}_{\mu \nu \cdots }^{\theta }=\mathcal{O}_{\mu \nu \cdots
}^{\dagger }(-1)^{\delta _{\mu 0}}(-1)^{\delta _{\nu 0}}\cdots . 
\]
Writing (\ref{distance}) we have used translational invariance, which we
assume here. We also assume that the $\mathcal{L}$-theory $\tau _{\mathcal{L}%
}$ is reflection positive. Instead, Lorentz invariance is not assumed. The
normalization factor appearing in (\ref{distance}) will be explained later.

The infinitesimal squared distance $\mathrm{d}\ell ^{2}$ defines the metric $%
g_{IJ}$ in parameter space: 
\begin{equation}
g_{IJ}(\hat{x},\lambda )=2\pi ^{4}\hat{x}^{8}\langle \mathcal{O}_{I}(\hat{x}%
_{v})\hspace{0.01in}\mathcal{O}_{J}^{\theta }(0)\rangle ,  \label{metric}
\end{equation}
which is a Hermitian matrix. Apart from the normalization factor, this
formula agrees with Zamolodchikov's definition \cite{zamo}. We often work in
a real basis, namely a basis where the parameters $\mathrm{d}\lambda ^{I}$
are real, the operators $\mathcal{O}_{I}$ are Hermitian and the metric is
symmetric.

If $-d_{I}$ denotes the canonical dimension of the parameter $\mathrm{d}%
\lambda ^{I}$ in units of mass, we can write 
\begin{equation}
\langle \mathcal{O}_{I}(\hat{x}_{v})\hspace{0.01in}\mathcal{O}_{J}^{\theta
}(0)\rangle =\frac{1}{2\pi ^{4}}\frac{G_{IJ}(t,\hat{\lambda})}{\hat{x}%
^{8+d_{I}+d_{J}}},  \label{green}
\end{equation}
where $t=-\ln (\hat{x}\mu )$, $\mu $ is the renormalization scale and $\hat{%
\lambda}^{I}=\hat{x}^{-d_{I}}\lambda $. Hence, 
\begin{equation}
g_{IJ}=\hat{x}^{-d_{I}-d_{J}}G_{IJ}(t,\hat{\lambda})  \label{metr}
\end{equation}
and the infinitesimal distance can be expressed as 
\begin{equation}
\mathrm{d}\ell =\sqrt{\mathrm{d}\hat{\lambda}^{I}\hspace{0.01in}G_{IJ}(t,%
\hat{\lambda})\hspace{0.01in}\mathrm{d}\hat{\lambda}^{*J}},  \label{dl0}
\end{equation}
where $\mathrm{d}\hat{\lambda}^{I}=\hat{x}^{-d_{I}}\mathrm{d}\lambda ^{I}$.

The $\hat{x}$ -dependence of the metric can be regarded as a dependence on
the energy. Its meaning will be illustrated with explicit examples. In the
appendix we prove that the distance $\mathrm{d}\ell $ is
renormalization-group invariant. In particular, $\mathrm{d}\ell $ can be
written in the manifestly RG-invariant form 
\[
\mathrm{d}\ell =\sqrt{\mathrm{d}\hat{\lambda}^{I}(t)\hspace{0.01in}G_{IJ}(0,%
\hat{\lambda}(t))\hspace{0.01in}\mathrm{d}\hat{\lambda}^{*J}(t)},
\]
where $\hat{\lambda}^{I}(t)=\hat{x}^{-d_{I}}\lambda ^{I}(t)$, and $\lambda
^{I}(t)$ denote the running coupling constants.

\bigskip

By reflection positivity, $\mathrm{d}\ell $ is non-negative. Terms
proportional to the field equations do not contribute to the distance,
because they give contact terms in the two-point function $\langle \mathrm{d}%
\mathcal{L}(\hat{x}_{v})~\mathrm{d}\mathcal{L}(0)\rangle $, which are
negligible because $\hat{x}_{v}\neq 0$.

Now we define the finite distance between two $\mathcal{L}$-theories $\tau _{%
\mathcal{L}_{1}}$ and $\tau _{\mathcal{L}_{2}}$ at some energy scale $E$.
Let $\gamma _{12}$ denote a path in parameter space connecting $\mathcal{L}%
_{1}$ to $\mathcal{L}_{2}$, namely a curve 
\[
\gamma _{12}:\qquad \rho \in [0,1]\rightarrow \lambda ^{I}(\rho ), 
\]
where the values of the couplings are referred to the energy $E$ and $%
\lambda ^{I}(0)$ and $\lambda ^{I}(1)$ are such that $\mathcal{L}(\lambda
(0))=\mathcal{L}_{1}$ and $\mathcal{L}(\lambda (1))=\mathcal{L}_{2}$. The
length $\ell _{\gamma }(\tau _{\mathcal{L}_{1}},\tau _{\mathcal{L}_{2}})$ of 
$\gamma _{12}$ is defined as 
\[
\ell _{\gamma }(\tau _{\mathcal{L}_{1}},\tau _{\mathcal{L}%
_{2}})=\int_{\gamma _{12}}\mathrm{d}\ell , 
\]
and $\mathrm{d}\ell $ is calculated at $\hat{x}=1/E$. We can define the
distance $d_{\mathcal{L}}(\tau _{\mathcal{L}_{1}},\tau _{\mathcal{L}_{2}})$
between $\tau _{\mathcal{L}_{1}}$ and $\tau _{\mathcal{L}_{2}}$ as the
minimum of $\ell _{\gamma }(\tau _{\mathcal{L}_{1}},\tau _{\mathcal{L}_{2}})$
on the set of paths $\gamma _{12}$ that connect them: 
\[
d_{\mathcal{L}}(\tau _{\mathcal{L}_{1}},\tau _{\mathcal{L}%
_{2}})=\min_{\gamma _{12}}\int_{\gamma _{12}}\mathrm{d}\ell . 
\]

We define an equivalence relation between $\mathcal{L}$-theories stating
that two $\mathcal{L}$-theories are equivalent when they are separated by
zero distance. Standard arguments allow us to prove that this is indeed an
equivalence relation. Then, it is easy to prove that $d_{\mathcal{L}}(\tau _{%
\mathcal{L}_{1}},\tau _{\mathcal{L}_{2}})$ does satisfy the properties of a
distance, namely: $i$) it is positive-definite, and equal to zero if and
only if $\tau _{\mathcal{L}_{1}}$ is equivalent to $\tau _{\mathcal{L}_{2}}$%
; $ii$) it is symmetric and $iii$) it satisfies the triangle inequality.

We can also define the distance $d_{\mathcal{L}}(\tau _{\mathcal{L}};S)$
between an $\mathcal{L}$-theory $\tau _{\mathcal{L}}$ and a surface of $%
\mathcal{L}$-theories $S$, as the minimum of $d_{\mathcal{L}}(\tau _{%
\mathcal{L}},\tau _{\mathcal{L}}^{\prime })$ with respect to the set of
points $\tau _{\mathcal{L}}^{\prime }$ belonging to the surface: 
\[
d_{\mathcal{L}}(\tau _{\mathcal{L}};S)=\min_{\tau _{\mathcal{L}}^{\prime
}\in S}d_{\mathcal{L}}(\tau _{\mathcal{L}},\tau _{\mathcal{L}}^{\prime }). 
\]
In the applications we have in mind, $\tau _{\mathcal{L}}$ will be the
Lorentz-violating theory and $S$ will be the Lorentz surface.

The definitions we have just given have a number of properties that deserve
discussion and a detailed analysis. For example, (\ref{metric}) shows that
the metric, the distance $d_{\mathcal{L}}$ and the equivalence relation
between $\mathcal{L}$-theories do depend on the energy scale. This
dependence is expected, since two theories may be separated by different
distances at low and high energies. Consider free scalars of different
masses: they are equivalent at high energies, but not at low energies. Thus
their distance must tend to zero in the ultraviolet limit and have finite
values at any other scale.

However, the distance $d_{\mathcal{L}}$ also depends on several arbitrary
choices, such as reparametrizations of space and time, field redefinitions,
total derivatives, and so on. If the theory is Lorentz-violating the
distance (\ref{distance}) also depends on the time axis chosen to define the 
$\theta $-operation. Basically, the definition we have given is tied to the
Lagrangian used to formulate the theory. Physically equivalent theories
described by different Lagrangians may be separated by non-vanishing
distances $d_{\mathcal{L}}$. This is why we have spoken of distance between $%
\mathcal{L}$-theories, so far, and not distance between theories. A way to
remedy to this drawback is as follows.

Consider the space of theories as a fiber bundle, where the base manifold is
the set of physical theories $\tau $, and the fiber is the set of $\mathcal{L%
}$-theories $\tau _{\mathcal{L}}$ that correspond to the same physical
theory $\tau $. We may call it the \textit{Lagrangian bundle}. The distance $%
d_{\mathcal{L}}$ is a distance in the Lagrangian bundle, not a distance in
the base manifold. We can view each fiber as a surface in the bundle, and
define the true distance $d(\tau _{1},\tau _{2})$ between two theories $\tau
_{1}$ and $\tau _{2}$ as the distance between their fibers. This is not the
end of the story, however. Indeed, if we apply this definition literally,
namely calculate the minimum of the $d_{\mathcal{L}}$-distances between all
formulations $\tau _{\mathcal{L}_{1}}$ and $\tau _{\mathcal{L}_{2}}$
associated with $\tau _{1}$ and $\tau _{2}$, we get in general a trivial
result. Thus, the minimum must be calculated imposing suitable constraints,
which we discuss case by case. Since the distance depends on the energy, the
constraint should fix the units in which energies are measured, among the
other things. In the case we are mostly interested in, namely the distance
between a Lorentz-violating theory and a Lorentz-invariant one, the natural
constraint is to require that the latter be formulated in the usual
manifestly covariant form. We write 
\begin{equation}
d(\tau _{1},\tau _{2})=\min_{\mathcal{L}_{1},\mathcal{L}_{2}}{\!\!^{\prime
}\ }d_{\mathcal{L}}(\tau _{\mathcal{L}_{1}},\tau _{\mathcal{L}_{2}}),
\label{foss}
\end{equation}
where the prime is meant to remind us that the minimization is subject to
constraints.

The distance $d(\tau _{1},\tau _{2})$ does not satisfy the triangle
inequality, because it is a distance between surfaces, not a distance
between points. Yet, it is satisfactory for most of our purposes, and we
take it as our definition of distance between theories.

In some cases it may be preferable to choose a definite cross-section in the
bundle. This amounts to choose a set of conventions or prescriptions to
associate a particular formulation $\tau _{\mathcal{\bar{L}}}$ with each
physical theory $\tau $. Then the distance between two theories is just 
\begin{equation}
d(\tau _{1},\tau _{2})=d_{\mathcal{L}}(\tau _{\mathcal{\bar{L}}_{1}},\tau _{%
\mathcal{\bar{L}}_{2}}).  \label{crosssec}
\end{equation}
This definition does satisfy the triangle inequality, but the choice of $%
\tau _{\mathcal{\bar{L}}}$ may be arbitrary. It can be viewed as a
particular case of (\ref{foss}), where the constraint is the cross-section.

In the paper we study these issues in detail and discuss various ways and
prescriptions to remove the ambiguities associated with them. A
Lorentz-violating theory may remove some of these ambiguities by itself,
since it selects a preferred reference frame.

\section{Distance from the Lorentz surface}

\setcounter{equation}{0}

In this section we study the distance between a Lorentz-violating theory and
the Lorentz surface. Let us first recall a few general facts, before
applying them to our case. Consider a space described by real coordinates $%
x^{\mu }$, with a symmetric metric $g_{\mu \nu }(x)$. Assume that a surface $%
S$ is described by the equations 
\[
S^{i}(x)=0, 
\]
or, equivalently, by the map 
\[
u^{a}\longmapsto x^{\mu }(u). 
\]
Then the vectors 
\[
v_{a}^{\mu }=\frac{\partial x^{\mu }(u)}{\partial u^{a}} 
\]
are tangent to the surface. Differentiating $S^{i}(x(u))=0$ we obtain the
normal vectors 
\begin{equation}
n_{\mu }^{i}=\frac{\partial S^{i}}{\partial x^{\mu }}(x(u)),  \label{nv}
\end{equation}
which indeed satisfy $n_{\mu }^{i}v_{a}^{\mu }=0$ for every $i$ and $a$.

A vector can be projected onto its component normal to $S$ by means of the
projector 
\[
P_{\nu }^{\mu }=\delta _{\nu }^{\mu }-v_{a}^{\mu }h^{ab}v_{b}^{\rho }g_{\rho
\nu }, 
\]
where the matrix $h^{ab}$ is the inverse of $g_{\mu \nu }v_{a}^{\mu
}v_{b}^{\nu }$. It is easy to check that $v_{a}^{\mu }g_{\mu \rho }P_{\nu
}^{\rho }=0$ and $P_{\alpha }^{\mu }P_{\nu }^{\alpha }=P_{\nu }^{\mu }$.

The distance $d(\tau ;S)$ from a point $\tau $ to the surface $S$ is defined
as the distance from $\tau $ to the closest point on the surface: 
\[
d(\tau ;S)=\min_{\gamma _{\tau ;S}}\int_{\gamma _{\tau ;S}}\sqrt{\mathrm{d}%
x^{\mu }g_{\mu \nu }\mathrm{d}x^{\nu }}=\int_{\bar{\gamma}_{\tau ;S}}\sqrt{%
\mathrm{d}x^{\mu }g_{\mu \nu }\mathrm{d}x^{\nu }}, 
\]
where $\gamma _{\tau ;S}$ is any path from $\tau $ to the surface and $\bar{%
\gamma}_{\tau ;S}$ is the shortest path from $\tau $ to the surface.

Let $\sigma $ denote the endpoint of $\bar{\gamma}_{\tau ;S}$ on the
surface. We can easily show that $\bar{\gamma}_{\tau ;S}$ intersects the
surface orthogonally to it. First observe that given a point $\tau ^{\prime
} $ on $\bar{\gamma}_{\tau ;S}$, the shortest path $\bar{\gamma}_{\tau
^{\prime };S}$ from $\tau ^{\prime }$ to the surface is precisely the
portion of $\bar{\gamma}_{\tau ;S}$ connecting $\tau ^{\prime }$ to $\sigma $%
. Indeed, if it were not, we could use $\bar{\gamma}_{\tau ^{\prime };S}$ to
build a path from $\tau $ to the surface shorter than $\bar{\gamma}_{\tau
;S} $.

Now, consider a point $\tau ^{\prime }$ infinitesimally close to $S$. We can
vary the endpoints of the infinitesimal straight paths connecting $\tau
^{\prime }$ to $S$ adding tangent vectors to $\mathrm{d}x^{\mu }/\mathrm{d}s$%
. The distance from $\tau ^{\prime }$ to $S$ reads 
\[
\mathrm{d}\ell ^{\prime }=\min_{c}\mathrm{d}s\sqrt{\left( \frac{\mathrm{d}%
x^{\mu }}{\mathrm{d}s}+c^{a}v_{a}^{\mu }\right) g_{\mu \nu }\left( \frac{%
\mathrm{d}x^{\nu }}{\mathrm{d}s}+c^{b}v_{b}^{\nu }\right) }. 
\]
Minimizing with respect to the constants $c^{a}$ we find

\begin{equation}
\mathrm{d}\ell ^{\prime }=\sqrt{\mathrm{d}x^{\mu }P_{\mu }^{\rho }g_{\rho
\sigma }P_{\nu }^{\sigma }\mathrm{d}x^{\nu }},  \label{fr}
\end{equation}
which is the infinitesimal distance calculated along the normal vector $%
P_{\rho }^{\mu }\mathrm{d}x^{\rho }$. Thus, the path $\bar{\gamma}_{\tau ;S}$
intersects $S$ orthogonally to it in $\sigma $, as claimed.

We also see that when $\tau $ is infinitesimally close to $S$ we can
calculate its distance from the surface simply using the formula 
\begin{equation}
\mathrm{d}\ell =\sqrt{\mathrm{d}x^{\mu }\gamma _{\mu \nu }\mathrm{d}x^{\nu }}
\label{dlr}
\end{equation}
with the ``reduced'' metric 
\begin{equation}
\gamma _{\mu \nu }=P_{\mu }^{\rho }g_{\rho \sigma }P_{\nu }^{\sigma }=g_{\mu
\nu }-g_{\mu \rho }v_{a}^{\rho }h^{ab}v_{b}^{\sigma }g_{\sigma \nu }.
\label{gr}
\end{equation}

\bigskip

Now we apply these arguments to define the distance between a quantum field
theory and the Lorentz surface. We first assume that the theory is
infinitesimally close to the Lorentz surface.

Consider a Lorentz-invariant theory defined by a Lagrangian $\mathcal{L}%
_{LI}(\lambda )$ with parameters $\lambda ^{a}$. Write the Lorentz-violating
theory as 
\[
\mathcal{L}_{LI}+\sum_{i}\zeta ^{i}\mathcal{O}_{i}^{LV}. 
\]
We work in a basis where the parameters $\zeta ^{i}$ are real and the
operators are Hermitian. There exists no unambiguous definition of
Lorentz-violating operators $\mathcal{O}_{i}^{LV}$, since Lorentz-invariant
terms $\mathcal{O}_{a}^{LI}$ can always be added to them. The parameters $%
\zeta ^{i}$ move away from the Lorentz surface, but not necessarily
orthogonally to it. Thus, we have to consider a more general perturbation
that includes displacements $\xi ^{a}=\mathrm{d}\lambda ^{a}$ tangent to the
Lorentz surface. We write 
\begin{equation}
\mathcal{L}_{LI}+\sum_{i}\zeta ^{i}\mathcal{O}_{i}^{LV}=\mathcal{L}%
_{LI}^{\prime }+\sum_{i}\zeta ^{i}\mathcal{O}_{i}^{LV}+\sum_{a}\xi ^{a}%
\mathcal{O}_{a}^{LI}.  \label{lpert}
\end{equation}

Let $\mathrm{d}\lambda ^{I}=(\zeta ^{i},\xi ^{a})$, $I=(i,a)$ denote the
coordinates in parameter space. The metric (\ref{metric}) reads 
\[
g_{IJ}=\left( 
\begin{tabular}{cc}
$g_{ij}$ & $g_{ib}$ \\ 
$g_{aj}$ & $g_{ab}$%
\end{tabular}
\right) =2\pi ^{4}\hat{x}^{8}\left( 
\begin{tabular}{cc}
$\langle \mathcal{O}_{i}^{LV}$\hspace{0.01in}$\mathcal{O}_{j}^{LV}\rangle $
& $\langle \mathcal{O}_{i}^{LV}$\hspace{0.01in}$\mathcal{O}_{b}^{LI}\rangle $
\\ 
$\langle \mathcal{O}_{a}^{LI}$\hspace{0.01in}$\mathcal{O}_{j}^{LV}\rangle $
& $\langle \mathcal{O}_{a}^{LI}$\hspace{0.01in}$\mathcal{O}_{b}^{LI}\rangle $%
\end{tabular}
\right) . 
\]

The vectors orthogonal to the surface can be worked out as explained above.
The Lorentz surface is described by the equations 
\[
\zeta ^{i}=0 
\]
and the most general normal vector can be written in the form 
\[
n^{I}=(\zeta ^{i},-h^{ab}g_{bj}\zeta ^{j}), 
\]
where the matrix $h^{ab}$ denotes the inverse of $g_{ab}$, $%
h^{ac}g_{cb}=\delta _{b}^{a}$. Indeed, lowering the index $I$, we can
immediately prove that $n^{I}$ is a linear combination of the normal vectors
(\ref{nv}), which here simply read 
\[
n_{J}^{i}=\delta _{J}^{i}=(\delta _{j}^{i},0). 
\]
Explicitly, 
\[
n_{I}=g_{IJ}n^{J}=(\gamma _{ij}\zeta ^{j},0), 
\]
where 
\[
\gamma _{ij}=g_{ij}-g_{ia}h^{ab}g_{bj} 
\]
is the reduced metric.

Now, assume that a theory (\ref{lpert}) is given, by which we mean that the $%
\zeta ^{i}$'s and the parameters $\lambda ^{a}$ of $\mathcal{L}_{LI}$ are
known by experimental measurements. The theory (\ref{lpert}) is our point $%
\tau $ in parameter space. If the $\zeta ^{i}$'s are infinitesimal, the
geodesic $\bar{\gamma}_{\tau ;S}$ can be approximated by 
\[
\bar{\gamma}_{\tau ;S}:\qquad \rho \in [0,1]\rightarrow \rho (\zeta
^{i},-h^{ab}g_{bj}\zeta ^{j}).
\]
The motion along $\bar{\gamma}_{\tau ;S}$ can be illustrated by the
``Lagrangian'' 
\begin{equation}
\mathcal{L}(\rho )=\mathcal{L}_{LI}^{\prime }+\rho \sum_{i}\zeta ^{i}%
\mathcal{O}_{i}^{LV}-\rho \sum_{a}h^{ab}g_{bj}\zeta ^{j}\mathcal{O}_{a}^{LI},
\label{lrho}
\end{equation}
where 
\[
\mathcal{L}_{LI}^{\prime }=\mathcal{L}_{LI}+\sum_{a}h^{ab}g_{bj}\zeta ^{j}%
\mathcal{O}_{a}^{LI}.
\]
The endpoint $\tau $ of the geodesic is $\mathcal{L}(1)=\mathcal{L}%
_{LI}+\sum_{i}\zeta ^{i}\mathcal{O}_{i}^{LV}$, namely the theory that fits
experimental observations. Instead, $\mathcal{L}(0)=\mathcal{L}_{LI}^{\prime
}$ identifies the endpoint $\sigma $ on the surface $S$. According to (\ref
{dlr}) and (\ref{gr}), the distance from $\tau $ to the surface is then 
\begin{equation}
d_{\tau ;S}=\sqrt{\zeta ^{i}\gamma _{ij}\zeta ^{j}}=\sqrt{\zeta
^{i}g_{ij}\zeta ^{j}-\zeta ^{i}g_{ia}h^{ab}g_{bj}\zeta ^{j}}.  \label{dQS}
\end{equation}
This result (\ref{dQS}) can also be obtained minimizing 
\[
\mathrm{d}\ell =\sqrt{\zeta ^{i}g_{ij}\zeta ^{j}+2\zeta ^{i}g_{ia}\xi
^{a}+\xi ^{a}g_{ab}\xi ^{b}}
\]
with respect to the tangent displacements $\xi ^{a}$.

Some remarks are in order. The Lagrangian (\ref{lrho}) is non-local, in
general, since it contains the metric $g_{IJ}$, which is determined by
correlation functions. However, $\mathcal{L}(\rho )$ should not be viewed as
the Lagrangian of a true quantum field theory, but just a formula to
describe the operations on parameter space that determine geodesics and
distances. Only the theory that matches experimental observations, which is $%
\mathcal{L}(1)$ here, needs to have a standard local form.

In terms of Green functions the infinitesimal distance between the theory
and the Lorentz-invariant surface is 
\begin{equation}
d_{L}(t,\hat{\zeta})=\sqrt{\hat{\zeta}^{i}\gamma _{ij}(t,\hat{\lambda})\hat{%
\zeta}^{*j}},  \label{dlz}
\end{equation}
where we have switched back to a generic, non-real basis, and 
\begin{equation}
\gamma _{ij}(t,\hat{\lambda})=G_{ij}(t,\hat{\lambda})-G_{ia}(t,\hat{\lambda}%
)H^{ab}(t,\hat{\lambda})G_{bj}(t,\hat{\lambda})  \label{redm}
\end{equation}
is the reduced metric, $H^{ab}(t,\hat{\lambda})$ being the inverse matrix of 
$G_{ab}(t,\hat{\lambda})$. Observe that all correlations functions appearing
in (\ref{redm}) are calculated in the Lorentz invariant theory. RG\
invariance extends to $d_{L}(t,\hat{\zeta})$ (see the appendix), the
manifestly RG-invariant formula being 
\[
d_{L}(t,\hat{\zeta})=\sqrt{\hat{\zeta}^{i}(t)\hspace{0.01in}\gamma _{ij}(0,%
\hat{\lambda}(t))\hspace{0.01in}\hat{\zeta}^{*j}(t)}=d_{L}(0,\hat{\zeta}%
(t)). 
\]

We have been working at the level of $\mathcal{L}$-theories here. Later we
discuss unwanted dependencies, such as those on coordinate
reparametrizations, and the minimization with respect to them.

If the point $\tau $ is not infinitesimally close to the surface $S$, the
distance between $\tau $ and $S$ is the length of the geodesic path that
connects $\tau $ to the surface and hits the surface orthogonally.

\paragraph{Approximations}

The infinitesimal distance (\ref{dlz}) is sufficient for most practical
applications, because experimental bounds ensure that the parameters of the
Lorentz violation are very small.

When the Lorentz invariant theory is weakly coupled we can set its
dimensionless couplings to zero, to a first approximation. For example, in
low-energy QED we can neglect the fine structure constant and evaluate the
correlation functions, the metric $\gamma _{ij}$ and the distance (\ref{dlz}%
) in the free-field limit. In this case, if we switch the masses off, the
metric $\gamma _{ij}$ is just a constant, otherwise it depends on $\hat{m}=m%
\hat{x}$. In section 6 we study also theories that include Lorentz-violating
higher-dimensional operators, multiplied by inverse powers of some scale $%
\Lambda _{L}$. There the distance depends also on $\hat{\Lambda}_{L}=\Lambda
_{L}\hat{x}$.

\section{Simple examples}

\setcounter{equation}{0}

In this section we discuss the simplest examples of infinitesimal and finite
distances, namely free relativistic massive and massless fields and
spontaneously broken theories. We show how the distance depends on the
energy scale and discuss the meaning of this dependence. We work in the
Euclidean framework.

\begin{figure}[t]
\centering{\includegraphics[width=0.3\textwidth]{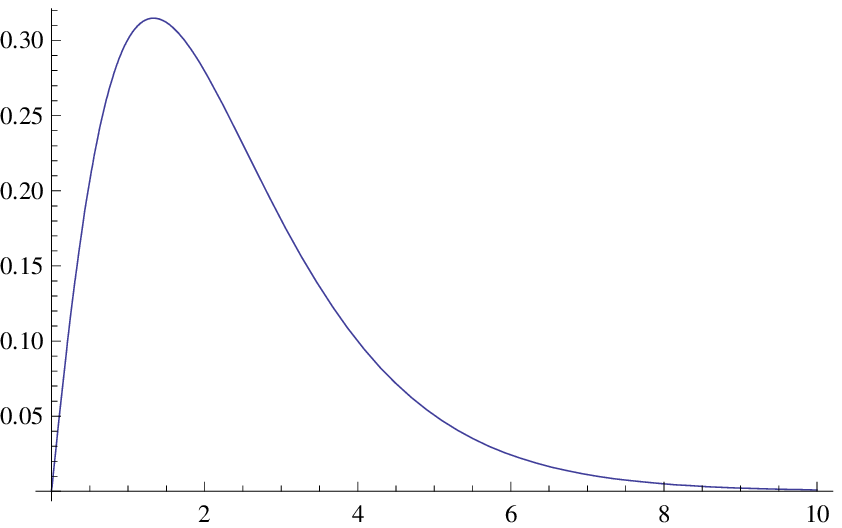} %
\includegraphics[width=0.3\textwidth]{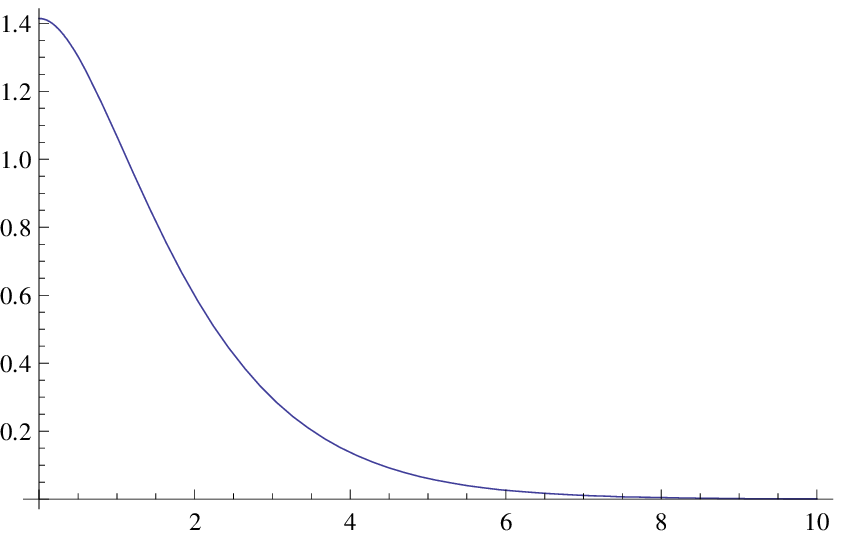} \includegraphics[width=0.3%
\textwidth]{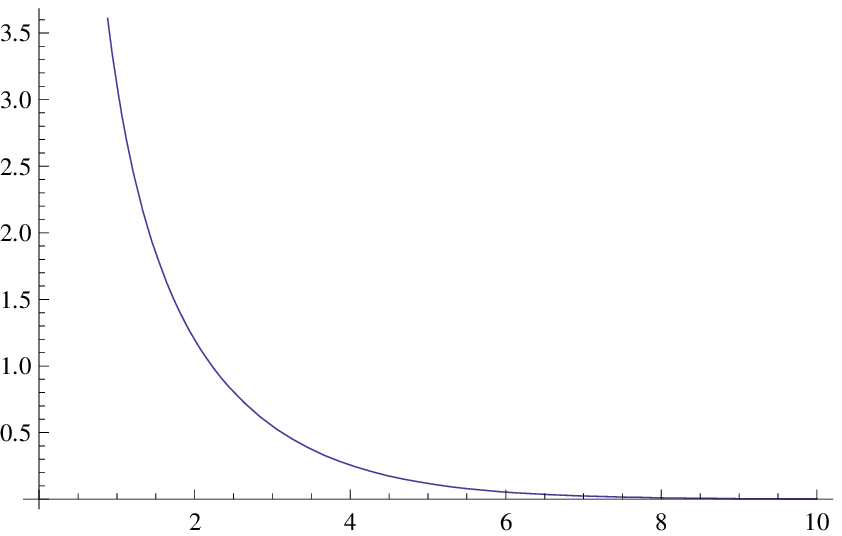}}
\caption{Plot of $\mathrm{d}{\ell }/\mathrm{d}u$ for massive scalars,
fermions and vectors, respectively.}
\label{dlf}
\end{figure}

We begin from the relativistic scalar field 
\[
\mathcal{L}_{s}=\frac{1}{2}(\partial _{\mu }\varphi )^{2}+\frac{m^{2}}{2}%
\varphi ^{2}
\]
and perturb the mass, that is to say we consider 
\[
\mathrm{d}\mathcal{L}_{s}=m\mathrm{d}m\hspace{0.01in}\varphi ^{2}. 
\]
Using the definition (\ref{distance}) the infinitesimal distance $\mathrm{d}%
\ell $ reads 
\begin{equation}
\mathrm{d}\ell =\sqrt{2\pi ^{4}\hat{x}^{8}\langle \mathrm{d}\mathcal{L}_{s}(%
\hat{x}_{v})\mathrm{\hspace{0.01in}d}\mathcal{L}_{s}(0)\rangle }=\frac{1}{2}%
u^{2}K_{1}(u)\mathrm{d}u,  \label{formula1}
\end{equation}
where $u=$ $m\hat{x}$ and $K_{n}$ denotes the modified Bessel function of
the second kind. The shape of $\mathrm{d}\ell /\mathrm{d}u$ is shown in Fig. 
\ref{dlf}, where it is also compared with analogue shapes for fermions and
vector fields.

The distance between two massive theories with $m_{2}>m_{1}$ is 
\begin{equation}
d(m_{2},m_{1})=\int_{m_{1}}^{m_{2}}\frac{\mathrm{d}\ell }{\mathrm{d}m}%
\mathrm{d}m=\frac{1}{2}\int_{m_{1}\hat{x}}^{m_{2}\hat{x}}u^{2}K_{1}(u)%
\mathrm{d}u.  \label{ds}
\end{equation}
Studying $d(m_{2},m_{1})$ as a function of $\hat{x}$, we note that

1) when both masses are non-vanishing, $m_{2}>m_{1}>0$, the distance is
different from zero for $\hat{x}\neq 0$, and tends to zero both in the
infrared limit $\hat{x}\rightarrow \infty $ and in the ultraviolet limit $%
\hat{x}\rightarrow 0$;

2) the distance between a massive theory $m_{2}=m>0$ and the massless theory 
$m_{1}=0$ tends to zero in the ultraviolet limit and to one in the infrared
limit. In particular, 
\begin{equation}
\left. d(m,0)\right| _{\hat{x}\rightarrow \infty }=1.  \label{norma}
\end{equation}

These behaviors are expected. Indeed, in the ultraviolet limit masses become
negligible, so a massive theory becomes equivalent to a massless one. In the
infrared limit massive theories become empty. However, a massless theory
remains non-empty in the infrared. There the distance between a massive
theory and a massless one tends to a non-vanishing constant. We have
normalized the distance to make this constant equal to one for one scalar
field.

Now we repeat the exercise for massive fermions, with 
\[
\mathcal{L}_{f}=\bar{\psi}(\slashed{\partial}+m)\psi ,\qquad \mathrm{d}%
\mathcal{L}_{f}=\mathrm{d}m\bar{\psi}\psi . 
\]
We get 
\begin{equation}
\mathrm{d}\ell =\pi ^{2}\hat{x}^{4}\sqrt{2\langle \mathrm{d}\mathcal{L}_{f}(%
\hat{x}_{v})\hspace{0.01in}\mathrm{d}\mathcal{L}_{f}(0)\rangle }=\frac{u^{2}%
\mathrm{d}u}{\sqrt{2}}\sqrt{K_{2}^{2}(u)-K_{1}^{2}(u)}.  \label{df}
\end{equation}
Numerically, we find 
\begin{equation}
\left. d(m,0)\right| _{\hat{x}\rightarrow \infty }=2.911.  \label{df0}
\end{equation}

Third, we consider massive vector fields, with 
\begin{equation}
\mathcal{L}_{v}=\frac{1}{4}W_{\mu \nu }^{2}+\frac{m^{2}}{2}W_{\mu
}^{2},\qquad \mathrm{d}\mathcal{L}_{v}=m\mathrm{d}mW_{\mu }^{2},
\label{procam}
\end{equation}
where $W_{\mu \nu }\equiv \partial _{\mu }W_{\nu }-\partial _{\nu }W_{\mu }$%
. We have 
\[
\mathrm{d}\ell =2\pi ^{2}\hat{x}^{4}m\mathrm{d}m\sqrt{\left( \langle W_{\mu
}(\hat{x}_{v})\hspace{0.01in}W_{\nu }(0)\rangle \right) ^{2}},\qquad \langle
W_{\mu }(x)\hspace{0.01in}W_{\nu }(0)\rangle =\left( \delta _{\mu \nu }-%
\frac{\partial _{\mu }\partial _{\nu }}{m^{2}}\right) \left( \frac{m}{4\pi
^{2}x}K_{1}^{2}(xm)\right) , 
\]
where $x=\sqrt{x_{1}^{2}+x_{2}^{2}+x_{3}^{2}+x_{4}^{2}}$. The explicit
formula of $\mathrm{d}\ell $ is an involved expression containing several
Bessel functions, which we do not write here. It tends to infinity in the
ultraviolet limit, which is expected, since a massive vector is singular
there. Precisely, 
\[
\left. \mathrm{d}\ell \right| _{u\rightarrow 0}\sim 2\sqrt{3}\frac{\mathrm{d}%
u}{u}. 
\]
In the same limit the distance $d(m_{2},m_{1})$ tends to a finite constant
if $m_{2}>m_{1}>0$: 
\[
\left. d(m_{2},m_{1})\right| _{\hat{x}\rightarrow 0}=2\sqrt{3}\ln \frac{m_{2}%
}{m_{1}}. 
\]
It tends to infinity with a logarithmic singularity when one mass tends to
zero: 
\[
\left. d(m_{2},m_{1})\right| _{m_{1}\rightarrow 0}\sim -2\sqrt{3}\ln (m_{1}%
\hat{x}). 
\]
Finally, it tends to zero in the infrared limit for $m_{2}>m_{1}>0$.

If a symmetry is spontaneously broken by the non-vanishing vacuum
expectation value $v$ of some scalar field $\varphi $, then the distance
between the symmetry-violating theory and the symmetric surface is
proportional to $\hat{v}=v\hat{x}=v/E$ and correctly tends to zero for
energies $E\gg v$. For example, consider the $\varphi ^{4}$-theory 
\[
\mathcal{L}=\frac{1}{2}(\partial _{\mu }\varphi )^{2}+\frac{\lambda }{4!}%
\varphi ^{4}-\frac{m^{2}}{2}\varphi ^{2}. 
\]
Expanding the scalar field around its expectation value $v$ we get 
\[
\mathcal{L}(\lambda ,v)=\frac{1}{2}(\partial _{\mu }\eta )^{2}+\frac{\lambda 
}{4!}\eta ^{4}+\frac{\lambda v}{3!}\eta ^{3}+\frac{\lambda v^{2}}{3!}\eta
^{2}\equiv \mathcal{L}(\lambda )+\mathrm{d}\mathcal{L}(\lambda ,v). 
\]
For $v$ small the distance has the form 
\[
d_{v}=\hat{v}(t)f(\lambda (t)), 
\]
where $\hat{v}(t)=v(t)\hat{x}$ and $\lambda (t)$ are the running couplings,
and $f$ is some function. When the energy $E=1/\hat{x}$ is much larger than $%
v$ (but smaller than the dynamical scale $\mu $, so that the perturbative
expansion in $\lambda $ is still meaningful\footnote{%
The sole purpose of this assumption is to ensure that the function $f$
remains bounded.}), the distance tends to zero and the symmetry is restored.
At higher energies the theory behaves like an ordinary $\varphi ^{4}$-theory.

\section{Marginal Lorentz-violating deformations of massless free
relativistic fields}

\setcounter{equation}{0}

In this section we calculate the distance in the case of the most general
marginal CPT-invariant Lorentz-violating deformations of massless
relativistic free fields. We study the effects of minimizations with respect
to tangent and reparametrization-displacements. We also use our results to
discuss a number of conventions.

In Minkowskian notation, we write the total Lagrangian 
\[
\mathcal{L}_{\text{LI}}+\mathrm{d}\mathcal{L}_{\text{LV}}, 
\]
as the sum of the Lorentz-invariant free Lagrangian 
\begin{equation}
\mathcal{L}_{\text{LI}}=\frac{1}{2}(\partial _{\mu }\varphi _{I})(\partial
^{\mu }\varphi _{I})+\frac{i}{2}\bar{\psi}_{A}\accentset{\leftrightarrow}{%
\partial }{\!\!\!\!\hspace{1.1pt}\slash}\psi _{A}-\frac{1}{4}F_{\mu \nu
}^{G}F_{{}}^{G\hspace{0.01in}\mu \nu }  \label{llli}
\end{equation}
and the Lorentz-violating perturbation 
\begin{equation}
\mathrm{d}\mathcal{L}_{\text{LV}}=\frac{1}{2}\mathrm{d}\epsilon _{IJ}^{\mu
\nu }(\partial _{\mu }\varphi _{I})(\partial _{\nu }\varphi _{J})+\frac{i}{2}%
\mathrm{d}c_{AB}^{\mu \nu }\bar{\psi}_{A}\gamma _{\mu }\accentset{%
\leftrightarrow }{\partial }_{\nu }\psi _{B}-\frac{1}{4}\mathrm{d}%
k_{GH}^{\mu \nu \rho \sigma }F_{\mu \nu }^{G}F_{\rho \sigma }^{H}.
\label{lllv}
\end{equation}
For the moment, we do not assume particular conditions on the coefficients $%
\mathrm{d}\epsilon _{IJ}^{\mu \nu }$, $\mathrm{d}c_{AB}^{\mu \nu }$ and $%
\mathrm{d}k_{GH}^{\mu \nu \rho \sigma }$, other than their obvious symmetry
and Hermiticity properties. Summations over repeated indices are understood.
The fermions are assumed to be chiral, although we do not need to specify
the chirality of each $\psi _{A}$.

Applying definition (\ref{distance}) we easily find the infinitesimal
squared distance 
\begin{equation}
\mathrm{d}\ell ^{2}=\sum_{\mu \nu =0}^{3}(\mathrm{d}\bar{\epsilon}_{IJ}^{\mu
\nu })(\mathrm{d}\epsilon _{IJ}^{\mu \nu })+2\sum_{\mu \nu =0}^{3}|\mathrm{d}%
c_{sAB}^{\mu \nu }|^{2}+3\sum_{\mu \nu =0}^{3}|\mathrm{d}\bar{c}_{aAB}^{\mu
\nu }|^{2}-\frac{1}{2}|\mathrm{trd}c_{AB}|^{2}+\sum_{\mu \nu \rho \sigma
=0}^{3}(\mathrm{d}k_{GH}^{\mu \nu \rho \sigma })^{2},  \label{margi}
\end{equation}
where 
\begin{eqnarray*}
\mathrm{d}c_{sAB}^{\mu \nu } &=&\frac{1}{2}(\mathrm{d}c_{AB}^{\mu \nu }+%
\mathrm{d}c_{AB}^{\nu \mu }),\qquad \mathrm{d}\bar{c}_{aAB}^{\mu \nu }=\frac{%
3^{-(\delta _{\mu 0}+\delta _{\nu 0})/2}}{2i}(\mathrm{d}c_{AB}^{\mu \nu }-%
\mathrm{d}c_{AB}^{\nu \mu }), \\
\mathrm{trd}c_{AB} &=&(\mathrm{d}c_{AB})_{\hspace{0.05in}\mu }^{\mu },\qquad
\qquad \qquad \mathrm{d}\bar{\epsilon}_{IJ}^{\mu \nu }=\frac{3^{\delta _{\mu
0}+\delta _{\nu 0}}}{4}\mathrm{d}\epsilon _{IJ}^{\mu \nu }.
\end{eqnarray*}
When the sum over space- and time-indices is written explicitly we mean that
indices are contracted with the Euclidean metric. It is easy to verify that
the two chiralities give the same contributions, which is why we do not need
to distinguish them explicitly.

Now we consider tangent and reparametrization-displacements. We minimize $%
\mathrm{d}\ell ^{2}$ with respect to each of them separately, and then
together. The tangent displacements are the Lorentz-invariant contributions
to $\mathrm{d}\mathcal{L}_{\text{LV}}$, namely 
\[
\frac{1}{2}\mathrm{d}\eta _{IJ}(\partial _{\mu }\varphi _{I})(\partial ^{\mu
}\varphi _{J})+\frac{i}{2}\mathrm{d}\tau _{AB}(\bar{\psi}_{A}%
\accentset{\leftrightarrow}{\partial }{\!\!\!\!\hspace{1.1pt}\slash}\psi
_{B})-\frac{1}{4}\mathrm{d}\zeta _{GH}(F_{\mu \nu }^{G}F^{\mu \nu H})-\frac{1%
}{8}\mathrm{d}\tilde{\zeta}_{GH}\varepsilon ^{\mu \nu \rho \sigma }(F_{\mu
\nu }^{G}F_{\rho \sigma }^{H}). 
\]
These displacements do not move us away from the Lorentz surface. Minimizing
(\ref{margi}) we obtain the conditions 
\begin{equation}
(\mathrm{d}\bar{\epsilon}_{IJ})_{\hspace{0.06in}\mu }^{\mu }=0,\qquad (%
\mathrm{d}k_{GH})_{\hspace{0.05in}\hspace{0.05in}\mu \nu }^{\mu \nu
}=0,\qquad \varepsilon _{\mu \nu \rho \sigma }\mathrm{d}k_{GH}^{\mu \nu \rho
\sigma }=0.  \label{tre}
\end{equation}
Observe that no conditions on the fermionic parameters $\mathrm{d}%
c_{AB}^{\mu \nu }$ are generated. The reason is that the Dirac Lagrangian is
proportional to its own field equations.

Infinitesimal reparametrizations of the form 
\begin{equation}
x^{\prime \hspace{0.01in}\mu }=x^{\mu }+\mathrm{d}a_{\nu }^{\hspace{0.05in}%
\mu }\hspace{0.01in}x^{\nu },\qquad A_{\mu }^{\prime }=A_{\mu }-\mathrm{d}%
a_{\mu }^{\hspace{0.05in}\nu }\hspace{0.01in}A_{\nu },  \label{repp}
\end{equation}
generate another noticeable subsector of $\mathrm{d}\mathcal{L}_{\text{LV}}$%
. Minimizing with respect to these displacements we obtain the relations 
\begin{equation}
\mathrm{d}C^{\mu \nu }-\frac{g^{\mu \nu }}{4}(\mathrm{d}C)_{\hspace{0.06in}%
\alpha }^{\alpha }=0,\qquad (\mathrm{d}B)_{\hspace{0.06in}\mu }^{\mu }=0,
\label{one}
\end{equation}
where we have defined 
\[
\mathrm{d}C^{\mu \nu }\equiv \mathrm{d}c_{AA}^{\mu \nu }+\mathrm{d}B^{\mu
\nu },\qquad \mathrm{d}B^{\mu \nu }\equiv \mathrm{d}\bar{\epsilon}_{II}^{\mu
\nu }+2(\mathrm{d}k_{GG})_{\hspace{0.09in}\hspace{0.09in}\alpha }^{\mu
\alpha \nu }. 
\]

Finally, if we minimize with respect to spinor reparametrizations 
\begin{equation}
\psi _{A}\rightarrow \psi _{A}+i\omega _{AB}^{\mu \nu }\sigma _{\mu \nu
}\psi _{B},  \label{spinpar}
\end{equation}
we obtain that the fermionic parameters must be symmetric, 
\begin{equation}
\mathrm{d}c_{AB}^{\mu \nu }=\mathrm{d}c_{AB}^{\nu \mu }.  \label{due}
\end{equation}

When we minimize with respect of all classes of displacements altogether, we
obviously obtain both (\ref{tre}), (\ref{one}) and (\ref{due}). These
conditions are Lorentz covariant in the absence of scalar fields. Note that
the first two formulas of (\ref{tre}) imply the second of (\ref{one}).

\bigskip

Now we further study the meaning of the minimization with respect to the
various kinds of displacements in simple examples. Consider $N$
Lorentz-violating rotation-preserving massless scalars, and take the
perturbed Lagrangian 
\begin{equation}
\mathcal{L}_{2s}+\mathrm{d}\mathcal{L}_{2s}=\frac{1}{2}\sum_{I}(\partial
_{\mu }\varphi _{I})(\partial ^{\mu }\varphi _{I})-\sum_{I}\mathrm{d}%
\epsilon _{I}(\partial _{i}\varphi _{I})^{2}.  \label{l2s}
\end{equation}
The na\"{i}ve squared distance between the Lorentz violating theory and the
Lorentz invariant one reads 
\[
\mathrm{d}\ell ^{2}=2\pi ^{4}\hat{x}^{8}\langle \mathrm{d}\mathcal{L}_{2s}(%
\hat{x}_{v})\mathrm{\hspace{0.01in}d}\mathcal{L}_{2s}(0)\rangle =3\sum_{I}(%
\mathrm{d}\epsilon _{I})^{2}. 
\]
However, if all $\mathrm{d}\epsilon _{I}$'s are equal the theory can be
still written in a manifestly Lorentz invariant form rescaling the space
coordinates. Thus, we write $\mathrm{d}\epsilon _{I}=\mathrm{d}\tilde{%
\epsilon}_{I}+\mathrm{d}a$ and minimize $\mathrm{d}\ell $ with respect to $%
\mathrm{d}a$. We obtain 
\[
\mathrm{d}\ell _{r}^{2}=\sum_{IJ}(\mathrm{d}\tilde{\epsilon}_{I})\gamma
_{IJ}(\mathrm{d}\tilde{\epsilon}_{J})=\frac{3}{N}\sum_{I<J}(\mathrm{d}\tilde{%
\epsilon}_{I}-\mathrm{d}\tilde{\epsilon}_{J})^{2}. 
\]
The entries of the reduced metric $\gamma _{IJ}$ are $3(1-1/N)$ on the
diagonal and $-3/N$ elsewhere. At the minimum the Lagrangian reads 
\[
\mathcal{L}_{2s}+\mathrm{d}\mathcal{L}_{2s}=\frac{1}{2}\sum_{I}(\partial
_{\mu }\varphi _{I})(\partial ^{\mu }\varphi _{I})-\sum_{I}\mathrm{d}\tilde{%
\epsilon}_{I}(\partial _{i}\varphi _{I})^{2}+\frac{1}{N}\sum_{IJ}\mathrm{d}%
\tilde{\epsilon}_{I}(\partial _{i}\varphi _{J})^{2}, 
\]
and satisfies the relations (\ref{one}), which just read $\sum_{I}\mathrm{d}%
\epsilon _{I}=0$ in this particular case.

Now, take two fields and assume that one is massive and the other one is
not: 
\begin{equation}
\mathcal{L}_{2s}+\mathrm{d}\mathcal{L}_{2s}=\frac{1}{2}\sum_{I=1}^{2}(%
\partial _{\mu }\varphi _{I})(\partial ^{\mu }\varphi _{I})-\frac{m^{2}}{2}%
\varphi _{1}^{2}-\sum_{I=1}^{2}\mathrm{d}\epsilon _{I}(\partial _{i}\varphi
_{I})^{2}.  \label{blao}
\end{equation}
Then we have a squared distance of the form 
\begin{equation}
\mathrm{d}\ell ^{2}=A(\mathrm{d}\epsilon _{1})^{2}+B(\mathrm{d}\epsilon
_{2})^{2},  \label{blag}
\end{equation}
with $A\neq B$. The minimization with respect to $\mathrm{d}a$ no longer
satisfies relations (\ref{one}), since now $\mathrm{d}\epsilon _{1}+\mathrm{d%
}\epsilon _{2}\neq 0$. Moreover, it introduces considerable complicacies,
because the functions $A$ and $B$ depend on the energy scale, among the
other things.

For this reason, sometimes it may not be convenient to minimize with respect
to re\-pa\-ra\-me\-tri\-za\-tion-displacements. Then the ambiguities
associated with reparametrizations can be eliminated by means of a
prescription. This corresponds to define the distance choosing a cross
section of the Lagrangian bundle, as shown in formula (\ref{crosssec}).

One example is to impose relations (\ref{one}) by default. This prescription
can be adopted in the most general Lorentz-violating theory, also when the
parameters $\mathrm{d}\epsilon $, $\mathrm{d}c$ and $\mathrm{d}k$ are not
infinitesimal.

If scalar fields are present, this prescription is not Lorentz covariant.
When we consider theories that are infinitesimally close to
Lorentz-invariant ones, we may want to adopt alternative Lorentz covariant
prescriptions. One example is to set the trace of fermion coefficients to
zero: 
\begin{equation}
\mathrm{d}c_{AA}^{\mu \nu }=0.  \label{kosteconv}
\end{equation}
This condition is sufficient to remove the ambiguity, while analogous
conditions on the scalar or vector coefficients remove only part of it.

It is always advisable to minimize with respect to tangent displacements,
since they correspond to movements on the Lorentz surface. However,
sometimes we may want to adopt prescriptions also for tangent displacements.
Doing so, we are not really calculating the distance from the
Lorentz-violating theory to the Lorentz surface, but the distance from the
Lorentz-violating theory to a particular point on the surface. Examples of
Lorentz covariant prescriptions for tangent displacements are 
\begin{equation}
(\mathrm{d}k_{GH})_{\quad \mu \nu }^{\mu \nu }=0,\qquad \varepsilon _{\mu
\nu \rho \sigma }\mathrm{d}k_{GH}^{\mu \nu \rho \sigma }=0,\qquad (\mathrm{d}%
c_{AB})_{\hspace{0.05in}\mu }^{\mu }=0,\qquad (\mathrm{d}\epsilon _{IJ})_{%
\hspace{0.05in}\mu }^{\mu }=0.  \label{kosteprescr}
\end{equation}
The first two are also found from the minimization, but the other two are
not.

\section{Distance in Lorentz-violating QED}

\setcounter{equation}{0}

Now we calculate the distance between Lorentz violating theories and the
Lorentz surface. Again, we assume, for simplicity, that CPT\ is preserved.
First we calculate $\mathrm{d}\ell $ in the low-energy sector of
Lorentz-violating QED. This means that we include the Lagrangian terms that
are renormalizable by ordinary power counting. Later we consider the QED\
subsector of the Lorentz-violating Standard Model of \cite{lvsm,noH}, which
includes higher-dimensional operators that are renormalizable by weighted
power counting. We also comment on the contributions of CPT-violating terms.

The low-energy Lagrangian of Lorentz-violating QED is $\mathcal{L}_{\text{LI}%
}+\mathrm{d}\mathcal{L}_{\text{LV}}$, where 
\begin{eqnarray*}
\mathcal{L}_{\text{LI}} &=&-\frac{1}{4}F^{\mu \nu }F_{\mu \nu }+\frac{i}{2}%
\bar{\psi}\accentset{\leftrightarrow}{\slashed{D}}{\psi }-m\bar{\psi}\psi ,
\\
\mathrm{d}\mathcal{L}_{\text{LV}} &=&-\frac{1}{4}(k_{F})_{\mu \nu \rho
\sigma }F^{\mu \nu }F^{\rho \sigma }+\frac{1}{2}\bar{\psi}\left( ic^{\mu \nu
}\gamma _{\mu }\accentset{\leftrightarrow}{D}_{\nu }+id^{\mu \nu }\gamma
_{5}\gamma _{\mu }\accentset{\leftrightarrow}{D}_{\nu }-H_{\mu \nu }\sigma
^{\mu \nu }\right) \psi ,
\end{eqnarray*}
where $D_{\mu }=\partial _{\mu }+ieA_{\mu }$ is the covariant derivative.
The parameters $k_{F}$ and $H_{\mu \nu }$ satisfy the symmetry properties 
\[
(k_{F})_{\mu \nu \rho \sigma }=(k_{F})_{\rho \sigma \mu \nu }=-(k_{F})_{\nu
\mu \rho \sigma },\qquad H_{\nu \mu }=-H_{\mu \nu }.
\]
For the moment, we do not assume other conditions on the parameters.

Since the Lorentz violating parameters are bound to have very small values,
to a first approximation we can neglect the fine structure constant and work
around the free-field limit of QED.

The distance does depend on the energy. At energies much smaller than the
electron mass $m_{e}$ the electron contribution is negligible and we can
work in the pure photon sector. At energies much greater than $m_{e}$ we can
work in the massless limit. In both cases we can use the formulas of the
previous section, possibly adding the contributions of relevant operators.

We minimize with respect to tangent displacements, and evaluate the distance
both before and after the minimization with respect to the
reparametrization-displacements (\ref{repp}).

\paragraph{Photon sector}

In the photon sector formula (\ref{margi}) gives 
\[
\mathrm{d}\ell _{\gamma }^{2}=\sum_{\mu \nu \rho \sigma =0}^{3}(k_{F}^{\mu
\nu \rho \sigma })^{2}. 
\]
Minimizing with respect to tangent displacements we obtain conditions (\ref
{tre}), which here read 
\begin{equation}
(k_{F})_{\quad \mu \nu }^{\mu \nu }=0,\qquad \varepsilon _{\mu \nu \rho
\sigma }k_{F}^{\mu \nu \rho \sigma }=0.  \label{convk}
\end{equation}
These relations leave 19 independent entries, out of 21. It is common \cite
{kmw} to express the surviving entries in terms of three traceless symmetric
matrices $\tilde{k}_{e+}$, $\tilde{k}_{e-}$ and $\tilde{k}_{o-}$, an
antisymmetric matrix $\tilde{k}_{o+}$ and a scalar $\tilde{k}_{\mathrm{tr}}$%
. After some manipulations we find 
\begin{equation}
\mathrm{d}\ell _{\gamma }^{2}=6\tilde{k}_{\mathrm{tr}}^{2}+2\sum_{ij}\left[ (%
\tilde{k}_{e-}^{ij})^{2}+(\tilde{k}_{o-}^{ij})^{2}+(\tilde{k}%
_{e+}^{ij})^{2}+(\tilde{k}_{o+}^{ij})^{2}\right] .  \label{dl2g}
\end{equation}
Three coefficients can be eliminated, because $\tilde{k}_{e+}$, $\tilde{k}%
_{e-}$ and $\tilde{k}_{o-}$ are traceless. The tabulated quantities \cite
{datatables} in the standard Sun-centered inertial reference frame are $(%
\tilde{k}^{11}-\tilde{k}^{22})_{e+,e-,o-}$ and $\tilde{k}_{e+,e-,o-}^{33}$.
Rewriting the result in terms of these, we have 
\begin{eqnarray}
\mathrm{d}\ell _{\gamma }^{2} &=&(\tilde{k}_{e-}^{11}-\tilde{k}%
_{e-}^{22})^{2}+(\tilde{k}_{e+}^{11}-\tilde{k}_{e+}^{22})^{2}+(\tilde{k}%
_{o-}^{11}-\tilde{k}_{o-}^{22})^{2}+4\sum_{i<j}\left( (\tilde{k}%
_{e-}^{ij})^{2}+(\tilde{k}_{e+}^{ij})^{2}+(\tilde{k}_{o-}^{ij})^{2}+(\tilde{k%
}_{o+}^{ij})^{2}\right)  \nonumber \\
&&+3(\tilde{k}_{e-}^{33})^{2}+3(\tilde{k}_{e+}^{33})^{2}+3(\tilde{k}%
_{o-}^{33})^{2}+6\tilde{k}_{\mathrm{tr}}^{2}.  \label{dlgamma}
\end{eqnarray}

We can now maximize \eqref{dlgamma} using the maximal sensitivities of ref. 
\cite{datatables}, Table III. The main contribution comes from the parameter
that is measured with the smallest precision, which is $\tilde{k}_{o+}^{12}$%
. We obtain 
\begin{equation}
\mathrm{d}\ell _{\gamma }\sim 2\tilde{k}_{o+}^{12}\leqslant 2\times 10^{-13}.
\label{fnd}
\end{equation}
If we include the CPT-violating contributions we obtain something like 
\[
\mathrm{d}\ell _{\gamma \text{CPT}}\lesssim 2\times 10^{-13}\sqrt{1+f\times
10^{-58}(\widehat{x}\hspace{0.01in}\mathrm{GeV})^{2}}, 
\]
where $f$ is a calculable numerical factor of order 1. This formula shows
that on the basis of present knowledge possible CPT-violating contributions
can be neglected for wavelengths smaller than 10$^{-3}$ light years.

Formula (\ref{fnd}) is the result obtained before minimizing with respect to
the reparametrizations (\ref{repp}). Such a minimization gives, from (\ref
{one}), the additional condition 
\begin{equation}
\tilde{k}^{\mu \nu }=0,  \label{fet}
\end{equation}
where 
\[
\tilde{k}^{\mu \nu }\equiv (k_{F})_{\hspace{0.05in}\hspace{0.05in}\hspace{%
0.06in}\alpha }^{\mu \alpha \nu }.
\]
It is easy to show that an alternative form of (\ref{dl2g}) is 
\begin{equation}
\mathrm{d}\ell _{\gamma }^{2}=2\sum_{\mu ,\nu =0}^{3}(\tilde{k}\,^{\mu \nu
})^{2}+2\sum_{ij}\left[ (\tilde{k}_{e+}^{ij})^{2}+(\tilde{k}%
_{o-}^{ij})^{2}\right] .  \label{fit}
\end{equation}
Then equation (\ref{fet}) gives 
\begin{equation}
\mathrm{d}\ell _{\gamma }\leqslant 6\times 10^{-32}.  \label{fnd2}
\end{equation}

This value is much smaller than (\ref{fnd}), because reparametrizations
allow us to cancel out all nonbirefringent parameters. This can be done only
in the absence of other particles.

\paragraph{Electron sector}

Now we consider the fermionic sector in the massless limit, where we can use
the formulas of the previous section, provided we add the contributions of
relevant operators.

Formula (\ref{margi}) and the $H$-contribution give 
\begin{equation}
\mathrm{d}\ell _{e}^{2}=\sum_{\mu \nu =0}^{3}\left[ 4(c^{\mu \nu
})^{2}+4(d^{\mu \nu })^{2}+\widehat{x}^{2}(H^{\mu \nu })^{2}\right] -(c_{\mu
}^{\hspace{0.05in}\mu })^{2}-(d_{\mu }^{\hspace{0.05in}\mu })^{2},
\label{dle2}
\end{equation}
plus terms proportional to the antisymmetric parts of $c^{\mu \nu }$ and $%
d^{\mu \nu }$. Minimizing with respect to the spinor reparametrizations (\ref
{spinpar}) we get (\ref{due}), which tells us that $c$ and $d$ are symmetric
matrices. Moreover, the traces of $c$ and $d$ cancel out, so we assume that
they vanish.

Maximizing (\ref{dle2}) using the maximal sensitivities of ref. \cite
{datatables}, Table II, for the electron, which is to date the only particle
except the photon whose parameters are fully measured (in the CPT-even
sector), we get 
\begin{equation}
\mathrm{d}\ell _{e}\leqslant \sqrt{3\times 10^{-28}+5\times 10^{-39}(%
\widehat{x}\hspace{0.01in}\mathrm{{GeV})^{2}}}\mathrm{.}  \label{dle}
\end{equation}
Since the zero-mass approximation is valid for energies greater than $m_{e}$%
, this result holds for $\widehat{x}\lesssim 1/m_{e}$. At energy $m_{e}$ we
have 
\begin{equation}
\mathrm{d}\ell _{e}\leqslant 2\times 10^{-14}.  \label{dleme}
\end{equation}

\paragraph{Low-energy Lorentz-violating QED}

At energies much greater than $m_{e}$ we can work in the massless limit.
Formula (\ref{margi}) gives 
\begin{equation}
\mathrm{d}\ell _{\text{QED}}^{2}=\sum_{\mu \nu \rho \sigma
=0}^{3}(k_{F}^{\mu \nu \rho \sigma })^{2}+\sum_{\mu \nu =0}^{3}\left[
4(c^{\mu \nu })^{2}+4(d^{\mu \nu })^{2}+\widehat{x}^{2}(H^{\mu \nu
})^{2}\right] -(c_{\mu }^{\hspace{0.05in}\mu })^{2}-(d_{\mu }^{\hspace{0.05in%
}\mu })^{2},  \label{dly}
\end{equation}
plus contributions proportional to the antisymmetric parts of $c^{\mu \nu }$
and $d^{\mu \nu }$. As before, the traces of the matrices $c$ and $d$ do not
contribute to the distance, so we assume that they vanish. Moreover,
minimizing with respect to tangent-displacements and spinor
reparametrizations, we get again (\ref{convk}) and that the matrices $c$ and 
$d$ are symmetric. In the end, the conditions we find coincide with the
currently adopted conventions \cite{colladay,qed}.

The distance is just $\mathrm{d}\ell _{\text{QED}}=\sqrt{\mathrm{d}\ell
_{\gamma }^{2}+\mathrm{d}\ell _{e}^{2}}$, where $\mathrm{d}\ell _{\gamma }$
and $\mathrm{d}\ell _{e}$ are given by (\ref{fnd}) and (\ref{dleme}),
respectively. Since $\mathrm{d}\ell _{e}$ is smaller than $\mathrm{d}\ell
_{\gamma }$ by one order of magnitude, we get $\mathrm{d}\ell _{\text{QED}%
}\sim 2\times 10^{-13}$. We recall that this result holds for $\widehat{x}%
\lesssim 1/m_{e}$. For $\widehat{x}\gtrsim 1/m_{e}$ we have instead (\ref
{fnd}).

These are the numerical values that we obtain before minimizing with respect
to the re\-pa\-ra\-me\-tri\-za\-tions (\ref{repp}). When we do minimize with
respect to them, we get, from (\ref{one}), 
\begin{equation}
\tilde{k}^{\mu \nu }+c^{\mu \nu }=0.  \label{formulastar}
\end{equation}
We can derive this condition more directly as follows. Using (\ref{fit}), we
can write 
\begin{equation}
\mathrm{d}\ell _{\text{QED}}^{2}=2\sum_{\mu ,\nu =0}^{3}(\tilde{k}\,^{\mu
\nu })^{2}+2\sum_{ij}\left[ (\tilde{k}_{e+}^{ij})^{2}+(\tilde{k}%
_{o-}^{ij})^{2}\right] +\sum_{\mu \nu =0}^{3}\left[ 4(c^{\mu \nu
})^{2}+4(d^{\mu \nu })^{2}+\widehat{x}^{2}(H^{\mu \nu })^{2}\right] .
\label{intee}
\end{equation}
Under reparametrizations (\ref{repp}) $\tilde{k}^{\mu \nu }\rightarrow 
\tilde{k}^{\mu \nu }-2a^{\mu \nu }-g^{\mu \nu }a_{\rho }^{\hspace{0.05in}%
\rho }$. To preserve the first of (\ref{convk}), we take a traceless $a^{\mu
\nu }$. Then the birefringent quantities $\tilde{k}_{e+}$ and $\tilde{k}_{o-}
$ are invariant, while the polarization-independent quantities $\tilde{k}%
_{e-}$, $\tilde{k}_{o+}$ and $\tilde{k}_{\mathrm{tr}}$ are collected in the
tensor $\tilde{k}^{\mu \nu }$. In the electron sector $c^{\mu \nu
}\rightarrow c^{\mu \nu }-a^{\mu \nu }$ under (\ref{repp}), while $d^{\mu
\nu }$ and $H^{\mu \nu }$ are invariant. Making the replacements in (\ref
{intee}) and minimizing with respect to $a^{\mu \nu }$, we get 
\begin{equation}
\mathrm{d}\ell _{\text{QED}}^{2}=\sum_{\mu \nu =0}^{3}\left[ \frac{2}{3}(%
\tilde{k}^{\mu \nu }-2c^{\mu \nu })^{2}+4(d^{\mu \nu })^{2}+\widehat{x}%
^{2}(H^{\mu \nu })^{2}\right] +2(\tilde{k}_{e+}^{ij})^{2}+2(\tilde{k}%
_{o-}^{ij})^{2}.  \label{dL}
\end{equation}
We point out the appearance of the combination $\tilde{k}^{\mu \nu }-2c^{\mu
\nu }$, which is indeed the one on which physical processes depend \cite{exa}%
. Formula (\ref{dL}) agrees with (\ref{formulastar}), because only $\tilde{k}%
^{\mu \nu }+c^{\mu \nu }=0$ turns (\ref{intee}) into (\ref{dL}).

We can now use the maximal sensitivities reported in ref. \cite{datatables},
Tables II and III, to give an upper bound on $\mathrm{d}\ell _{\text{QED}}$.
The coefficients $\tilde{k}_{e+}$ and $\tilde{k}_{o-}$ are constrained to be
smaller than $10^{-32}$, so we can ignore them. Again the only relevant
terms are the ones that contain $\tilde{k}_{o+}^{12}$. We find 
\[
\mathrm{d}\ell _{\text{QED}}\leqslant \sqrt{10^{-26}+5\times 10^{-39}(%
\widehat{x}\hspace{0.01in}\mathrm{{GeV})^{2}}}.
\]
This formula holds for $\widehat{x}\lesssim 1/m_{e}$, where the second
contribution under the square root is negligible, so we get 
\begin{equation}
\mathrm{d}\ell _{\text{QED}}\leqslant \frac{2}{\sqrt{3}}\tilde{k}%
_{o+}^{12}\leqslant 10^{-13}.  \label{dlqed}
\end{equation}
For $\widehat{x}\gtrsim 1/m_{e}$ we have instead (\ref{fnd2}).

\paragraph{QED subsector of the high-energy-Lorentz-violating Standard Model}

Now we calculate the distance in the QED\ subsector of the Lorentz-violating
Standard Model of refs. \cite{lvsm,noH}. For simplicity, we assume that
rotations are preserved, besides parity and CPT, and concentrate on the
photon sector. We consider two choices for the total-derivative terms, and
compare the results we obtain. In the next section the dependence on
total-derivative terms is analysed in more detail.

We have seen that in general the distance can depend on the coordinate
parametrization. In the previous sections we have eliminated this ambiguity
minimizing with respect to re\-pa\-ra\-me\-tri\-za\-tion-displacements or
choosing some prescriptions. In some cases Lorentz-violating theories
eliminate the problem by themselves, because they already choose a preferred
reference frame.

For example, the Lorentz-violating Standard Model of ref.s \cite{lvsm,noH}
has a Lagrangian that contains higher space derivatives, to ensure
renormalizability by weighted power counting. However, that Lagrangian does
not contain higher time derivatives, to ensure perturbative unitarity. A
reparametrization (\ref{repp}) spoils this structure unless $\mathrm{d}%
a_{0}^{\hspace{0.05in}i}\hspace{0.01in}=0$. Moreover, it is assumed that
there exists a preferred frame where the theory is invariant under spatial
rotations. Then, if we want to preserve manifest rotational invariance we
must also have $\mathrm{d}a_{i}^{\hspace{0.05in}0}\hspace{0.01in}=0$ and $%
\mathrm{d}a_{i}^{\hspace{0.05in}j}\hspace{0.01in}=\delta _{j}^{i}\mathrm{d}a$%
. In the end, only time- and space-rescalings survive.

We study the Lagrangians $\mathcal{L}_{\text{LI}}+\mathrm{d}\mathcal{L}_{%
\text{LV}}$ and $\mathcal{L}_{\text{LI}}+\mathrm{d}\mathcal{L}_{\text{LV}%
}^{\prime }$, where 
\[
\mathcal{L}_{\text{LI}}=-\frac{1+\delta _{1}}{4}F_{\mu \nu }F^{\mu \nu } 
\]
and 
\begin{eqnarray}
\mathrm{d}\mathcal{L}_{\text{LV}} &=&-\frac{\delta _{2}}{4}(F^{ij})^{2}+%
\frac{\tau _{1}}{4\Lambda _{L}^{2}}F^{ij}\partial _{k}^{2}F^{ij}-\frac{\tau
_{0}}{4\Lambda _{L}^{4}}(\partial _{k}^{2}F^{ij})(\partial _{l}^{2}F^{ij}), 
\nonumber \\
\mathrm{d}\mathcal{L}_{\text{LV}}^{\prime } &=&-\frac{\delta _{2}}{4}%
(F^{ij})^{2}-\frac{\tau _{1}}{4\Lambda _{L}^{2}}(\partial _{k}F^{ij})^{2}-%
\frac{\tau _{0}}{4\Lambda _{L}^{4}}(\partial _{k}\partial _{l}F^{ij})^{2}.
\label{lhe}
\end{eqnarray}
A $\delta _{1}$-variation is a tangent displacement. Time- and
space-rescalings correspond to appropriate variations of $\delta _{1}$ and $%
\delta _{2}$. We minimize with respect to $\delta _{1}$ and keep $\delta
_{2} $ as an independent coupling.

The two distances we find are 
\begin{eqnarray}
\mathrm{d}\ell _{\text{QED}}^{2} &=&\frac{3}{2}\delta _{2}^{2}+60\frac{%
\delta _{2}\tau _{1}}{\hat{\Lambda}_{L}^{2}}+1200\frac{\delta _{2}\tau _{0}}{%
\hat{\Lambda}_{L}^{4}}+1260\frac{\tau _{1}^{2}}{\hat{\Lambda}_{L}^{4}}+76800%
\frac{\tau _{0}\tau _{1}}{\hat{\Lambda}_{L}^{6}}+1876800\frac{\tau _{0}^{2}}{%
\hat{\Lambda}_{L}^{8}},  \nonumber \\
\mathrm{d}\ell _{\text{QED}}^{\prime \hspace{0.01in}2} &=&\frac{3}{2}\delta
_{2}^{2}-24\frac{\delta _{2}\tau _{1}}{\hat{\Lambda}_{L}^{2}}+480\frac{%
\delta _{2}\tau _{0}}{\hat{\Lambda}_{L}^{4}}+384\frac{\tau _{1}^{2}}{\hat{%
\Lambda}_{L}^{4}}-3840\frac{\tau _{0}\tau _{1}}{\hat{\Lambda}_{L}^{6}}+505920%
\frac{\tau _{0}^{2}}{\hat{\Lambda}_{L}^{8}},  \label{didi}
\end{eqnarray}
respectively. We see that the two expressions have the same qualitative
features. Either of them (or any other Lagrangian differing from $\mathrm{d}%
\mathcal{L}_{\text{LV}}$ and $\mathrm{d}\mathcal{L}_{\text{LV}}^{\prime }$
by total-derivative terms) can be used to define the distance.

At present the ratio $\tau _{0}/\Lambda _{L}^{4}$ is constrained to be
smaller than $10^{-24}\mathrm{GeV}^{-4}$, while $|\tau _{1}|/\Lambda
_{L}^{2} $ is smaller than $10^{-21}\mathrm{GeV}^{-2}$ \cite{mewes}. With
these values and $\delta _{2}=-2\tilde{k}_{\mathrm{tr}}$, $|\delta
_{2}|\leqslant 2\times 10^{-14}$, we get from (\ref{didi}) the upper bounds 
\begin{eqnarray}
\mathrm{d}\ell _{\text{QED}}^{2} &\leqslant &6\times 10^{-28}\left(
1+2\times 10^{-6}q^{2}+4\times 10^{-8}q^{4}+10^{-13}q^{6}+3\times
10^{-15}q^{8}\right) ,  \nonumber \\
\mathrm{d}\ell _{\text{QED}}^{\prime \hspace{0.01in}2} &\leqslant &6\times
10^{-28}\left( 1+8\times 10^{-7}q^{2}+2\times 10^{-8}q^{4}+6\times
10^{-15}q^{6}+8\times 10^{-16}q^{8}\right) ,  \label{diere}
\end{eqnarray}
where $q=\hat{x}\mathrm{GeV}$. The formulas are valid as long as $\mathrm{d}%
\mathcal{L}_{\text{LV}}$ or $\mathrm{d}\mathcal{L}_{\text{LV}}^{\prime }$
are small, namely up to $10^{6}\mathrm{GeV}$. At this energy we find 
\begin{equation}
\mathrm{d}\ell _{\text{QED}}^{2}\leqslant 2\times 10^{6},\qquad \mathrm{d}%
\ell _{\text{QED}}^{\prime \hspace{0.01in}2}\leqslant 5\times 10^{5}.
\label{fis}
\end{equation}
Ignoring the $\tau _{0}$-term the resulting formulas are valid up to $10^{10}%
\mathrm{GeV}$, where they give 
\begin{equation}
\mathrm{d}\ell _{\text{QED}}^{2}\leqslant 13,\qquad \mathrm{d}\ell _{\text{%
QED}}^{\prime \hspace{0.01in}2}\leqslant 4\text{.}  \label{s}
\end{equation}

These distances are large because at present the bounds on the parameters of
higher-di\-men\-sio\-nal corrections are not so strong. In particular,
formula (\ref{fis}) reflects the bound on $\tau _{0}$, while formula (\ref{s}%
) reflects the bound on $\tau _{1}$.

Observe that the large numerical coefficients appearing in formulas (\ref
{diere}) can be turned into coefficients of order one if we incorporate a
factor 1/6 in $\Lambda _{L}$. This means that the Lorentz violation
introduced by higher-derivative terms starts to become important at energies 
$E\sim \Lambda _{Lr}\equiv \Lambda _{L}/6$, rather than $E\sim \Lambda _{L}$%
, as naively suggested by the Lagrangians (\ref{lhe}). In other physical
quantities the effects of higher-dimensional operators may be enhanced even
more.

CPT-violating quadratic corrections to the photon lagrangian carry odd
powers of momentum. Normally they are relevant operators by weighted power
counting, therefore in the ultraviolet limit their contributions to the
distance are negligible with respect to some prevailing CPT-even
contributions.

\section{Coordinate changes, total derivatives and other dependencies}

\setcounter{equation}{0}

\label{rescalings}In section 4 we illustrated how the metric and the
distance depend on the energy scale, and shown that such dependencies have
physically reasonable behaviors. However, the metric depends on several less
meaningful parameters, which may be introduced changing coordinates and
adding total derivatives. Precisely, it does depend on the Lagrangian used
to calculate it. In this section we illustrate these dependencies further
more.

Theories that are perturbatively equivalent can be described by Lagrangians
differing by total derivatives. Since the metric is based on the two-point
function of the Lagrangian perturbation $\mathrm{d}\mathcal{L}$, a
total-derivative perturbation can generate a non-vanishing distance. Let us
consider free massive scalar fields again, but now rescale them by some
factor $\mathrm{e}^{\Omega /2}$. We take the Lagrangian 
\[
\mathcal{L}_{\Omega }=\frac{\mathrm{e}^{\Omega }}{2}(\partial _{\mu }\varphi
)^{2}+\mathrm{e}^{\Omega }\frac{m^{2}}{2}\varphi ^{2}. 
\]
The difference $\mathcal{L}_{\Omega }-\mathcal{L}_{0}$ is not proportional
to the field equations, so it does contribute to the distance. Now the
two-parameter perturbation reads 
\[
\mathrm{d}\mathcal{L}_{\Omega }=\mathcal{L}_{\Omega }\mathrm{d}\Omega +%
\mathrm{e}^{\Omega }m\mathrm{d}m\hspace{0.01in}\varphi ^{2} 
\]
and the metric is 
\begin{equation}
\left( 
\begin{tabular}{cc}
$g_{mm}$ & $g_{m\Omega }$ \\ 
$g_{\Omega m}$ & $g_{\Omega \Omega }$%
\end{tabular}
\right) =2\pi ^{4}\hat{x}^{8}\left( 
\begin{tabular}{cc}
$\mathrm{e}^{2\Omega }m^{2}\langle \varphi ^{2}(\hat{x}_{v})\hspace{0.01in}%
\hspace{0.01in}\varphi ^{2}(0)\rangle $ & $\mathrm{e}^{\Omega }m\langle 
\hspace{0.01in}\varphi ^{2}(\hat{x}_{v})\hspace{0.01in}\mathcal{L}_{\Omega
}(0)\rangle $ \\ 
$\mathrm{e}^{\Omega }m\langle \mathcal{L}_{\Omega }(\hat{x}_{v})\hspace{%
0.01in}\hspace{0.01in}\varphi ^{2}(0)\rangle $ & $\langle \mathcal{L}%
_{\Omega }(\hat{x}_{v})\hspace{0.01in}\mathcal{L}_{\Omega }(0)\rangle $%
\end{tabular}
\right) .  \label{omega}
\end{equation}
In this simple case the metric is $\Omega $-independent, which is important
for the reason we explain below.

Consider an arbitrary trajectory $\Omega (m)$ from $m_{1}$ to $m_{2}$. The
distance calculated along the path $\Omega (m)$ is 
\begin{equation}
d[\Omega ]=\int_{m_{1}}^{m_{2}}\mathrm{d}m\sqrt{g_{mm}+2g_{m\Omega }\Omega
^{\prime }+g_{\Omega \Omega }\Omega ^{\prime \hspace{0.01in}2}}.
\label{dlam}
\end{equation}
Since theories with different values of $\Omega $ are physically equivalent,
the distance should not depend on the boundary values $\Omega (m_{1})$ and $%
\Omega (m_{2})$. A distance with such a property can be defined minimizing $%
d[\Omega ]$ with respect to the paths $\Omega (m)$, with free boundary
conditions. The minimization gives 
\begin{equation}
\Omega ^{\prime }=-\frac{g_{m\Omega }}{g_{\Omega \Omega }},  \label{lp}
\end{equation}
an orthogonality relation similar to the ones found before. Finally, the
distance between two massive theories reads 
\begin{equation}
d(m_{2},m_{1})=\int_{m_{1}}^{m_{2}}\mathrm{d}m\sqrt{g_{mm}-\frac{g_{m\Omega
}^{2}}{g_{\Omega \Omega }}}.  \label{gg}
\end{equation}

This result can be easily generalized to the case of several massive
scalars. If we take a unique overall rescaling factor $\mathrm{e}^{\Omega }$
the metric is still $\Omega $-independent. In that case, considering $m$ as
a vector, we can easily prove that the distance obeys the triangle
inequality 
\begin{equation}
d(m_{3},m_{2})+d(m_{2},m_{1})\geqslant d(m_{3},m_{1}).  \label{dd}
\end{equation}
Indeed, since the metric is $\Omega $-independent, the minimizing
trajectories $\Omega (m_{i},m_{j})$ connecting $m_{i}$ and $m_{j}$ can be
freely translated. We can use a translation to join the endpoints of $\Omega
(m_{3},m_{2})$ and $\Omega (m_{2},m_{1})$ at $m_{2}$, which gives a path $%
\bar{\Omega}(m)$ connecting $m_{1}$ with $m_{3}$. Thus, the right-hand side
of (\ref{dd}) is not greater than the left-hand side, since it is obtained
minimizing with respect to a set of paths that includes $\bar{\Omega}(m)$.

In the one-scalar case (\ref{gg}) gives an involved expression that we do
not report here, but obviously the distance is smaller than (\ref{ds}). For
example, the distance between a massive and a massless field in the infrared
limit turns out to be 
\begin{equation}
\left. d(m,0)\right| _{\hat{x}\rightarrow \infty }=0.482,  \label{blu}
\end{equation}
instead of 1.

Assume now that a problem (\ref{dlam}) is given, where, however, the metric
does depend on $\Omega $ (an explicit example is given below). Then the
minimization with respect to paths $\Omega (m)$ with free boundary
conditions gives the Euler equations plus the boundary conditions $%
g_{m\Omega }+g_{\Omega \Omega }\Omega ^{\prime }=0$ at $m_{1}$ and $m_{2}$.
The boundary conditions determine both $\Omega (m_{1})$ and $\Omega (m_{2})$%
, so it is impossible to paste the trajectories $\Omega (m_{3},m_{2})$ and $%
\Omega (m_{2},m_{1})$ at $m_{2}$ and prove the triangle inequality. If we
want to eliminate the $\Omega $-arbitrariness and keep the triangle
inequality we must choose a convention, for example demand that the kinetic
term $\dot{\varphi}^{2}/2$ be normalized to one. Then the correct results
are (\ref{ds}) and (\ref{norma}).

If we repeat the rescaling exercise for fermions, we still get (\ref{df})
and (\ref{df0}), since the fermion rescaling generates a perturbation
proportional to the field equations. In the case of massive vectors we find
results similar to the ones of the scalar case.

\bigskip

Now we study how the distance depends on the coordinate frame in more
detail. In formula (\ref{distance}) a time axis is chosen to apply
reflection positivity. If the theory is Lorentz invariant, and expressed in
manifestly Lorentz covariant form, the choice of time axis does not affect
the distance. However, a Lorentz invariant theory can also be expressed in a
form that is not manifestly Lorentz covariant. For example, we can make a
coordinate transformation 
\begin{equation}
x_{\mu }^{\prime }=(A^{-1})_{\mu \nu }x_{\nu },  \label{repra}
\end{equation}
$A$ being any real invertible matrix. Then the distance does depend on $A$.
Explicitly, let us consider a scalar field with Lagrangian 
\begin{equation}
\mathcal{L}_{B}=\frac{1}{2}B_{\mu \nu }\partial _{\mu }\varphi \partial
_{\nu }\varphi +\frac{m^{2}}{2}\varphi ^{2}.  \label{fra}
\end{equation}
This theory is equivalent to the relativistic scalar field after the
replacement (\ref{repra}) plus 
\begin{equation}
\varphi ^{\prime }(x^{\prime })=\sqrt{\det A}\hspace{0.01in}\varphi (x),
\label{pla}
\end{equation}
where $B=A^{2}$. We want to calculate the distance between two massive
theories in the reference frame (\ref{fra}). We have 
\[
\mathrm{d}\mathcal{L}_{B}(x)=m\mathrm{d}m\hspace{0.01in}\varphi ^{2}(x)=%
\frac{m\mathrm{d}m}{\det A}\hspace{0.01in}\varphi ^{\prime \hspace{0.01in}%
2}(x^{\prime }), 
\]
and in the primed frame we can use the formulas found above. We find 
\[
\mathrm{d}\ell =\pi ^{2}\hat{x}^{4}\sqrt{2\langle \mathrm{d}\mathcal{L}_{B}(%
\hat{x}_{v})~\mathrm{d}\mathcal{L}_{B}(0)\rangle }=\frac{2\pi ^{2}\hat{x}%
^{4}m\mathrm{d}m}{\det A}\hspace{0.01in}\langle \varphi ^{\prime }(\hat{x}%
_{v}^{\prime })~\varphi ^{\prime }(0)\rangle , 
\]
where 
\[
(\hat{x}_{v}^{\prime })_{\mu }=(A^{-1})_{\mu \nu }(\hat{x}_{v})_{\nu
}=(A^{-1})_{\mu 0}\hat{x},\qquad |\hat{x}_{v}^{\prime }|=\sqrt{(B^{-1})_{00}}%
\hat{x}. 
\]
We finally get 
\begin{equation}
\mathrm{d}\ell =\frac{u^{\prime \hspace{0.01in}2}\mathrm{d}u^{\prime }%
\hspace{0.01in}K_{1}\left( u^{\prime }\right) }{2(B^{-1})_{00}^{2}\sqrt{\det
B}}\hspace{0.01in},\qquad u^{\prime }=m\hat{x}\sqrt{(B^{-1})_{00}}.
\label{formula2}
\end{equation}
With respect to (\ref{formula1}), we just get rescaling factors, yet the
formula shows that the distance does depend on the choice of coordinate
frame. For example, the distance between a massive and a massless scalar
field in the infrared limit is 
\[
\left. d(m,0)\right| _{\hat{x}\rightarrow \infty }=\frac{1}{(B^{-1})_{00}^{2}%
\sqrt{\det B}}. 
\]
If we minimize with respect to $B_{\mu \nu }$ we find zero. Thus, in general
when we minimize with respect to the full set of Lagrangian formulations we
get a trivial result. Instead, as explained in section 2, the minimization
with respect to the Lagrangian formulation must be subject to constraints.
Typically, it is sufficient to fix the form of one theory. If one theory is
Lorentz invariant, it is sufficient to demand that it be formulated in a
manifestly covariant form.

Instead of treating $B_{\mu \nu }$ as fixed matrix, we can consider
deformations of both $B_{\mu \nu }$ and $m$. Using a matrix notation, we
then have 
\[
\mathrm{d}\mathcal{L}_{B}(x)=(\partial \varphi )^{T}\frac{\mathrm{d}B}{2}%
\partial \varphi (x)+m\mathrm{d}m\hspace{0.01in}\varphi ^{2}(x)=\frac{1}{%
\det A}\left[ (\partial ^{\prime }\varphi ^{\prime })^{T}A^{-1}\frac{\mathrm{%
d}B}{2}\hspace{0.01in}A^{-1}\partial ^{\prime }\varphi ^{\prime }+m\mathrm{d}%
m\hspace{0.01in}\varphi ^{\prime \hspace{0.01in}2}\right] ,
\]
where $T$ denotes transposition. After some straightforward manipulations we
obtain 
\begin{equation}
\mathrm{d}\ell ^{2}=\frac{u^{\prime \hspace{0.01in}4}}{16(B^{-1})_{00}^{4}%
\det B}\left\{ \mathrm{tr}\left[ \mathrm{d}B\hspace{0.01in}C\mathrm{d}B%
\hspace{0.01in}C\right] +4u^{\prime \hspace{0.01in}2}K_{2}^{\prime \hspace{%
0.01in}2}\mathrm{tr}[\mathrm{d}B\hspace{0.01in}\tilde{B}^{-1}]\frac{\mathrm{d%
}m}{m}+4u^{\prime \hspace{0.01in}2}K_{1}^{\prime \hspace{0.01in}2}\frac{(%
\mathrm{d}m)^{2}}{m^{2}}\right\} \hspace{0.01in},  \label{omica}
\end{equation}
where 
\[
\tilde{B}^{-1}=\frac{B^{-1}\mathds{1}_{00}B^{-1}}{(B^{-1})_{00}},\qquad
C=K_{2}^{\prime }B^{-1}-u^{\prime }K_{3}^{\prime }\tilde{B}^{-1},\qquad
K_{n}^{\prime }\equiv K_{n}(u^{\prime }),
\]
and $\mathds{1}_{00}$ is the matrix with entries $(\mathds{1}_{00})_{\mu \nu
}=\delta _{\mu 0}\delta _{\nu 0}$. Comparing (\ref{omica}) with (\ref{omega}%
), we see that now the metric does depend on $B$, while (\ref{omega}) did
not depend on $\Omega $.

We should now minimize with respect to $B$, namely solve the corresponding
Euler equations and boundary conditions. This is difficult to do, so here
for illustrative purposes we just minimize $\mathrm{d}\ell ^{2}$ with
respect to the reparametrization-displacements $\delta \mathrm{d}B_{\mu \nu
} $. We obtain 
\begin{equation}
\mathrm{d}\ell =\frac{u^{\prime \hspace{0.01in}3}}{2(B^{-1})_{00}^{2}\sqrt{%
\det B}}\frac{\mathrm{d}m}{m}\sqrt{K_{1}^{\prime \hspace{0.01in}2}-\frac{%
u^{\prime \hspace{0.01in}2}K_{2}^{\prime \hspace{0.01in}4}}{(K_{2}^{\prime
}-u^{\prime }K_{3}^{\prime }\hspace{0.01in})^{2}}}.  \label{formula3}
\end{equation}
The dependence on $B$ is even more involved than in (\ref{formula2}). Again,
the minimum with respect to $B$ gives a trivial result. Setting $\sqrt{\det B%
}=(B^{-1})_{00}=1$ we get 
\[
\left. d(m,0)\right| _{\hat{x}\rightarrow \infty }=0.466, 
\]
which is even smaller than (\ref{blu}).

To summarize, if we do not minimize with respect to $B$ the ambiguity in the
choice of $B$ survives. If we do minimize we get a trivial result. To obtain
a non-trivial result, we must minimize with a suitable constraint.
Generally, we lose the triangle inequality. In practice, after the
constrained minimization the distance has the properties of a distance
between surfaces, not the properties of a distance between points. Obviously
the distance between surfaces does not obey a simple triangle inequality.

Observe that in principle we should consider not just the linear
reparametrizations (\ref{repra}), but the most general curved
reparametrizations, and minimize (with constraints) with respect to them.
This is another reason why in several situations it may be more convenient
to remove the ambiguities associated with coordinate reparametrizations and
total derivatives by means of prescriptions.

At the same time, it is interesting to observe is that all definitions share
the same qualitative properties. For example, plotting (\ref{formula3}) we
see that it has the same shape as the first of Fig. \ref{dlf}. Despite its
unusual features, we do believe that the distance defined here has the good
properties to quantify the amount of symmetry violations.

\section{Conclusions}

\setcounter{equation}{0}

We have studied the distance between symmetry-violating quantum field
theories and the surface of symmetric theories. If a symmetry is known to be
violated in Nature, the distance measures how small, or large, the violation
is. If a symmetry, such as Lorentz symmetry, is not known to be violated in
Nature, then, using experimental bounds on the parameters of the violation,
the distance can be useful to quantify how precise the symmetry is at
present. We stress that although at the Lagrangian level an explicit
symmetry violation is in general described by a large number of independent
parameters, the symmetry violation \textit{per se} is actually governed by a
single quantity, such as the distance we have studied here.

Our results can be applied to any symmetry, and also to study the distance
between any pair or sets of theories. Here our main interest was to measure
the current precision of Lorentz symmetry. We have focused on the QED\
subsector, first at low energies and later including higher-dimensional
operators.

The distance has a number of interesting properties, but also unusual
features. For example, it depends on total derivatives and coordinate
parametrizations. In general, unwanted dependencies can be eliminated
minimizing with respect to the parameters associated with them, but
sometimes this procedure introduces more complicacies. Then it may be
preferable to choose prescriptions. We have shown that in the massless limit
of QED, the minimization gives constraints that coincide with the
conventions currently used in the literature.

\vskip 10truept \noindent {\Large \textbf{Acknowledgements}}

\vskip 5truept

D.Anselmi wishes to thank Xinmin Zhang and the Institute of High Energy
Physics of the Chinese Academy of Sciences, Beijing, for hospitality.
D.Anselmi is supported by the Chinese Academy of Sciences, grant No.
2010T2J01.

\vskip 20truept \noindent {\Large \textbf{Appendix: RG invariance of the
distance}}

\vskip 10truept

\renewcommand{\theequation}{A.\arabic{equation}} \setcounter{equation}{0}

In this appendix we prove that the infinitesimal distances (\ref{dl0}) and (%
\ref{dlz}) are renormalization-group invariant. Although strictly speaking
the argument we give applies to theories that are renormalizable by ordinary
power counting, it can be immediately generalized to theories that are
renormalizable by weighted power counting \cite{halat}, once the dimensions
of parameters and operators are replaced with their weights. Then $\mathrm{d}%
\ell ^{2}$ is invariant along the ``weighted'' RG flow \cite{halat}.

We start from (\ref{dl0}). The renormalization-group equations give, in
matrix notation, 
\[
G(t,\hat{\lambda})=z^{-1}(t)\hspace{0.01in}G(0,\hat{\lambda}(t))\hspace{%
0.01in}(z^{-1}(t))^{\dagger }. 
\]
The running renormalization constants read 
\[
z(t)=1+\sum_{k=1}^{\infty }(-1)^{k}\int_{0}^{t}\mathrm{d}t_{1}%
\int_{0}^{t_{1}}\mathrm{d}t_{2}\cdots \int_{0}^{t_{k-1}}\mathrm{d}t_{k}\
\gamma (t_{1})\ \cdots \ \gamma (t_{k-1})\ \gamma (t_{k}) 
\]
and $\gamma (t)$ stands for $\gamma (\hat{\lambda}(t))$, where 
\[
\gamma _{IJ}=Z_{IK}^{-1}\frac{\mathrm{d}Z_{KJ}}{\mathrm{d}\ln \mu },\qquad 
\mathcal{O}_{I\mathrm{B}}=Z_{IJ}\mathcal{O}_{J}, 
\]
the subscript B denoting bare quantities.

The running $\mathrm{d}\lambda $-parameters are $\mathrm{d}\hat{\lambda}(t)=%
\mathrm{d}\hat{\lambda}\hspace{0.01in}z^{-1}(t)$, so we can express the
infinitesimal length by means of the manifestly RG invariant formula 
\[
\mathrm{d}\ell =\sqrt{\mathrm{d}\hat{\lambda}^{I}(t)\hspace{0.01in}G_{IJ}(0,%
\hat{\lambda}(t))\hspace{0.01in}\mathrm{d}\hat{\lambda}^{J*}(t)}.
\]
Observe that the runnings of $\hat{\lambda}$ and $\mathrm{d}\hat{\lambda}$
are related by the formula $\beta _{\mathrm{d}\lambda }^{\hspace{0.05in}I}=%
\mathrm{d}\lambda ^{J}(\partial \beta ^{I}/\partial \lambda ^{J})$, where $%
\beta ^{I}$ is the beta function of $\lambda ^{I}$. This can be proved
shifting $\lambda $ by $\mathrm{d}\lambda $ in $\mu \mathrm{d}\lambda ^{I}/%
\mathrm{d}\mu =\beta ^{I}(\lambda )$.

The proof of RG-invariance can be extended to (\ref{dlz}), where the metric
is replaced by the reduced metric, obtained minimizing with respect to the
displacements $\xi ^{a}$ tangent to the Lorentz surface. First observe that
when the Lorentz-violating parameters $\zeta ^{i}$ identically vanish the
theory is consistently renormalizable, because it is Lorentz invariant. This
implies that the renormalization constants $Z_{ai}$ vanish, so $Z_{IJ}$ is
block triangular. Using this fact, it is easy to prove that 
\[
\gamma _{ij}(t,\hat{\alpha})=z_{ik}^{-1}(t)\hspace{0.01in}\gamma _{km}(0,%
\hat{\alpha}(t))\hspace{0.01in}z_{jm}^{-1*}(t), 
\]
so finally the distance (\ref{dlz}) depends only on the running couplings $%
\hat{\zeta}^{i}(t)=\hat{\zeta}^{j}z_{ji}^{-1}(t)$: 
\[
d_{L}(t,\hat{\zeta})=\sqrt{\hat{\zeta}^{i}\gamma _{ij}(t,\hat{\alpha})\hat{%
\zeta}^{*j}}=\sqrt{\hat{\zeta}^{i}(t)\hspace{0.01in}\gamma _{ij}(0,\hat{%
\alpha}(t))\hspace{0.01in}\hat{\zeta}^{*j}(t)}=d_{L}(0,\hat{\zeta}(t)). 
\]

\end{document}